\documentclass[a4paper,12pt]{article}
\pdfoutput=1

\usepackage{amsfonts,amsthm,amsmath,amssymb,graphicx,hyperref,color,youngtab}
\def\Xint#1{\mathchoice
   {\XXint\displaystyle\textstyle{#1}}%
   {\XXint\textstyle\scriptstyle{#1}}%
   {\XXint\scriptstyle\scriptscriptstyle{#1}}%
   {\XXint\scriptscriptstyle\scriptscriptstyle{#1}}%
   \!\int}
\def\XXint#1#2#3{{\setbox0=\hbox{$#1{#2#3}{\int}$}
     \vcenter{\hbox{$#2#3$}}\kern-.5\wd0}}

\def\dashint{\Xint-}

\newcommand{\be}{\begin{equation}}
\newcommand{\bea}{\begin{eqnarray}}
\newcommand{\ee}{\end{equation}}
\newcommand{\eea}{\end{eqnarray}}

\begin{document}

\makeatletter
\@addtoreset{equation}{section}
\makeatother
\renewcommand{\theequation}{\thesection.\arabic{equation}}

\rightline{}
\vspace{1.8truecm}

\vspace{15pt}


{\LARGE{  
\centerline{\bf Gauge Invariants, Correlators and Holography}
\centerline{\bf in Bosonic and Fermionic Tensor Models} 
}}  

\vskip.5cm 

\thispagestyle{empty}
    {\large \bf 
\centerline{Robert de Mello Koch\footnote{ {\tt robert@neo.phys.wits.ac.za}}, 
David Gossman\footnote{ {\tt dmgossman@gmail.com}} 
and  Laila Tribelhorn\footnote{ {\tt laila.tribelhorn@gmail.com}}}}

\vspace{.4cm}
\centerline{{\it National Institute for Theoretical Physics ,}}
\centerline{{\it School of Physics and Mandelstam Institute for Theoretical Physics,}}
\centerline{{\it University of Witwatersrand, Wits, 2050, } }
\centerline{{\it South Africa } }

\vspace{1.4truecm}

\thispagestyle{empty}

\centerline{\bf ABSTRACT}

\vskip.4cm 
Motivated by the close connection of tensor models to the SYK model, we use representation theory to
construct the complete set of gauge invariant observables for bosonic and fermionic tensor models.
Correlation functions of the gauge invariant operators in the free theory are computed exactly.
The gauge invariant operators close a ring.
The structure constants of the ring are described explicitly.
Finally, we construct a collective field theory description of the bosonic tensor model.

\setcounter{page}{0}
\setcounter{tocdepth}{2}

\newpage

\tableofcontents

\setcounter{footnote}{0}

\linespread{1.1}
\parskip 4pt

{}~
{}~

\section{Introduction}

The SYK model\cite{Kitaev,Sachdev:1992fk} may provide a simple solvable example of
holography\cite{Maldacena:1997re}, realized as an AdS/CFT 
duality - see \cite{Polchinski:2016xgd,Jevicki:2016bwu,Maldacena:2016hyu,Jevicki:2016ito}.
This expectation is motivated by the fact that the model develops an approximate conformal symmetry in the infrared.
Exact conformal symmetry is spontaneously and explicitly broken, leading to a pseudo-Goldstone mode.
This mode is responsible for the exponential growth of out of time ordered correlators, which saturates the 
chaos bound\cite{Maldacena:2015waa}. 
Saturating the bound is a strong hint that the model is dual to something close to Einstein gravity.
Much of the progress to date is possible because the large $N$ limit is dominated by a simple class of diagrams.
It is because these diagrams can be summed that the model is solvable, even at strong coupling.

The SYK model describes fermions interacting with all-to-all random interactions.
However, the large $N$ physics of the SYK model is identical to a tensor model, that has a
conventional large $N$ limit\cite{Witten:2016iux}.
The large $N$ limit of the tensor models is dominated by melonic 
graphs\cite{Gurau:2016lzk,Klebanov:2016xxf,Ferrari:2017ryl,Itoyama:2017emp}, which can be summed.
For interesting related work on holographic tensor models see \cite{Carrozza:2015adg}-\cite{Yoon:2017nig}.
Earlier work on tensor models includes \cite{Gurau:2016cjo}-\cite{Gurau:2010ba}.

The mechanism by which gravitational physics is manifested from a strongly coupled gauge theory remains
elusive.
The original CFT description has the field theory coupling as the loop expansion parameter.
On the other hand, the gravitational description that emerges at strong coupling, must have $1/N$ as the loop 
counting parameter.
This is a highly non-trivial hint into the structure of the holographic reorganization of the CFT. 
The collective field theory of Jevicki and Sakita\cite{collft} achieves exactly this: by formulating the theory in terms 
of gauge invariant variables, the resulting field theory explicitly has $1/N$ as the loop expansion parameter.
The reorganization of the dynamics is highly non-trivial, with non-linear collective dynamics being
induced by the Jacobian of the change of variables\cite{collft}.

It would be very attractive to apply the collective field theory method to CFTs and explore the resulting field theory.
In the case of a single matrix, this leads to a string field theory for the $c=1$ string\cite{Das:1990kaa}.
This is a beautiful example of how a quantum mechanical system can develop an extra dimension. 
In the much more interesting example of ${\cal N}=4$ super Yang-Mills theory\cite{Maldacena:1997re} (or even QCD)
the construction of collective field theory is frustrated by the fact that the space of gauge invariants (loop space) is
enormous and an explicit construction of the dynamics of gauge invariants looks hopeless.
It turns out that representation theory can provide a systematic approach towards the structure of loop space.
Indeed, the use of representation theory in the half-BPS sector\cite{Corley:2001zk} leads to a clear connection to 
free fermions\cite{Corley:2001zk,Berenstein:2004kk} and ultimately to a rather complete understanding of the 
mapping between the CFT operators and supergravity geometries\cite{Lin:2004nb}. 
This has been extended to more general bosonic sectors\cite{Balasubramanian:2004nb} -\cite{Kimura:2012hp}
and even for fermions and gauge fields\cite{deMelloKoch:2011vn,Koch:2012sf}.
These bases allow the computations of anomalous dimensions of heavy operators in ${\cal N}=4$ super 
Yang-Mills\cite{Koch:2011hb,Carlson:2011hy,deMelloKoch:2011ci,deMelloKoch:2012ck}
(in a large $N$ but non planar limit\cite{Balasubramanian:2001nh}) that are dual to excited giant 
gravitons\cite{de Mello Koch:2007uu}-\cite{Koch:2015pga}.
Up to now however, even with this improved understanding, it is not obvious how to build the collective field theory 
of these invariant variables.

Vector models are much simpler.
The space of invariants is spanned by a bilocal field and one can explicitly build the collective dynamics\cite{deMelloKoch:1996mj}.
In \cite{Das:2003vw} the idea that the bilocal fields provide a reconstruction of the bulk fields of the dual
higher spin gravity\cite{Klebanov:2002ja} was put forwards.
Using essentially kinematics \cite{Koch:2010cy}-\cite{Koch:2014aqa} developed a map between the space of 
bilocals and the dual gravity.
The bilocal description has also proved to be very useful for the SYK model 
itself\cite{Jevicki:2016bwu,Jevicki:2016ito,Das:2017pif},
as well as for descriptions of supersymmetric versions 
of SYK\cite{Fu:2016vas,Peng:2016mxj,Peng:2017spg,Murugan:2017eto}, 
in which case an elegant bilocal superspace formulation has been developed in \cite{Yoon:2017gut}.

One might hope that the case of tensor models is, in a sense, intermediate between the vector and matrix models.
It is possible that the space of gauge invariants is richer than that of vectors, but still not as complex as that of
matrices.
If this is the case, this may provide a useful lesson towards managing the loop space of multi-matrix models.
We explore this possibility in the present article.

Our basic goal is to construct the gauge invariants of both bosonic and fermionic tensor models.
For bosonic tensor (colored as well as non-colored) models,  the paper \cite{Geloun:2013kta} counted the 
gauge invariants uncovering a relationship with counting problems of branched covers of the 2-sphere.
The rank $d$ of the tensor is related to a number of branch points. 
Further, formulas for correlators of the tensor model invariants in a permutation basis were obtained.
Correlators in the permutation basis have been related to  the (Hurwitz) character calculus in
\cite{Mironov:2017aqv} (see also \cite{Koch:2010zza}).
A dual representation theory basis was developed in \cite{Diaz:2017kub}.
Our starting point reconsiders the representation theory basis for the bosonic tensor models, in a way that naturally 
allows an extension to fermionic tensor models. 
The basic ideas are explained in the next section, where we obtain counting formulas for the number of gauge invariant
operators in bosonic and fermionic tensor models.
The counting results for the bosons agree with results presented in \cite{Geloun:2013kta,Diaz:2017kub,will}.
The counting formulas for the fermions are new.
In section 3 we consider both the computations of the vacuum expectation values of our gauge invariant
operators, as well as two point functions of normal ordered gauge invariant operators.
These computations are performed exactly (i.e. to all orders in $1/N$), in the free theory.
In section 4 we describe the algebraic structure of the gauge invariants: they form a ring.
In section 5, we construct a collective field theory in terms of a subset of the gauge invariant variables in analogy to the 
construction for a single matrix model. We exhibit an emergent dimension and show that the Hamiltonian is local in this
new dimension. We reproduce large $N$ correlators of the tensor model quantum mechanics from the classical
collective field theory. Finally, in section 6 we conclude and mention some possible directions for further investigation.

\section{Construction of Gauge Invariant Operators}\label{CountConstruct}

In this section we simply want to count the number of gauge invariant operators that can be constructed, for both bosonic and fermionic tensor models. Once we have understood how to count the number of gauge invariants, a natural construction formula will be evident.

The fields that we consider are tensors, of rank $r$. We will denote the bosonic tensors by $\phi_{b_1 b_2 \cdots b_r}$ and the fermionic tensors by $\psi_{f_1 f_2 \cdots f_r}$. These fields transform in the fundamental of $G=U(N_1)\times U(N_2)\times\cdots\times U(N_r)$. 

Let $V_k$ denote the vector space carrying a copy of the fundamental representation of $U(N_k)$. Fields transforming in the fundamental of $G=U(N_1)\times U(N_2)\times\cdots\times U(N_r)$ belong to ${\cal V}\equiv V_1\times V_2\times \cdots\times V_r$. To build gauge invariants we will also need fields that transform in the anti-fundamental, denoted $\bar\phi^{b_1 b_2 \cdots b_r}$ and $\bar\psi^{f_1 f_2 \cdots f_r}$. Gauge invariants are then given by contracting corresponding upper and lower indices. The valid gauge invariants, built using two fields, are given by
\bea
\bar\phi^{b_1 b_2 \cdots b_r}\phi_{b_1 b_2 \cdots b_r}\qquad 
\bar\psi^{f_1 f_2 \cdots f_r}\psi_{f_1 f_2 \cdots f_r}
\eea
The operators that follow are not observables because they are not gauge invariant
\bea
\bar\phi^{b_1 b_2 \cdots b_r}\phi_{b_2 b_1 \cdots b_r}\qquad 
\bar\psi^{f_1 f_2 f_3\cdots f_r}\psi_{f_1 f_3 f_2 \cdots f_r}
\eea
A valid gauge invariant operator is only obtained if we contract corresponding indices of the tensors, since
the position of an index signifies which gauge group it belongs to.
To simplify the arguments that follow, we now specialize to rank 3 tensors $\phi_{ijk}$ or $\psi_{ijk}$ with 
$i=1,...,N_1$, $j=1,...,N_2$ and $k=1,...,N_3$.
The generalization to higher rank tensors is completely clear.
A comment is in order: the symmetry $U(N) \times U(N) \times U(N)$ can not be realized in any interacting theory, 
whose large $N$ expansion is dominated by melonic diagrams. 
The maximal symmetry in this case is only $U(N) \times U(N) \times O(N)$.

We will want to consider products of tensors to build the general gauge invariant operator.
Here is an example
\bea
\phi_{i_1j_1k_1}\phi_{i_2j_2k_2}\cdots\phi_{i_nj_nk_n}
\eea
This notation will quickly get out of hand, as the number of indices rapidly proliferates.
To avoid this, we will now use the notation first introduced in \cite{Corley:2001zk}.
The sleek notation uses a capital Roman letter to collect all of the little Roman letter indices, for example 
$I$ stands for $i_1,i_2,\cdots,i_n$.
We will also use a capital Greek letter to collect the tensors.
Thus, for example, we write
\bea
\Phi_{IJK}=\phi_{i_1j_1k_1}\phi_{i_2j_2k_2}\cdots\phi_{i_nj_nk_n}
\eea
Similarly
\bea
\Psi_{IJK}=\psi_{i_1j_1k_1}\psi_{i_2j_2k_2}\cdots\psi_{i_nj_nk_n}
\eea
These fields belong to ${\cal V}^{\otimes n}$.
There is a natural action of $S_n$ on ${\cal V}^{\otimes n}$ defined as follows: For any $\sigma\in S_n$ we have
\bea
\sigma\cdot\Phi_{IJK}\to \Phi_{\sigma (I)\sigma (J)\sigma (K)}
=\phi_{i_{\sigma(1)} j_{\sigma(1)} k_{\sigma(1)}}
\phi_{i_{\sigma(2)} j_{\sigma(2)} k_{\sigma(2)}}\cdots
\phi_{i_{\sigma(n)}j_{\sigma(n)}k_{\sigma(n)}}
\eea
We will sometimes call this the diagonal action of $S_n$ since each type of index, $i$, $j$ or $k$ is permuted
in exactly the same way.
We could also define an action of $S_n\times S_n\times S_n$ that acts independently on these three indices.
The notation distinguishing these two actions is
\bea
\sigma\cdot\Phi_{IJK}\to \Phi_{\sigma (I)\sigma (J)\sigma (K)}\qquad\sigma\in S_n
\eea
versus
\bea
\sigma_1\circ\sigma_2\circ\sigma_3\cdot\Phi_{IJK}\to \Phi_{\sigma_1 (I)\sigma_2 (J)\sigma_3 (K)}
\qquad \sigma_1\circ\sigma_2\circ\sigma_3\in S_n\times S_n\times S_n
\eea
Since the diagonal action swaps the tensors we have
\bea
\sigma\cdot\Phi_{IJK}=\Phi_{\sigma (I)\sigma (J)\sigma (K)}=\Phi_{IJK}\label{bos}
\eea
\bea
\sigma\cdot\Psi_{IJK}=\Psi_{\sigma (I)\sigma (J)\sigma (K)}={\rm sgn}(\sigma)\Psi_{IJK}\label{fer}
\eea
We know that swapping fermions costs a sign which is what the above equation captures.
In the last formula above ${\rm sgn}(\sigma)$ denotes the signature of the permutation $\sigma$.
For example, if $n=2$ and $\sigma=(12)$ we have
\bea
(12)\psi_{i_1 j_1 k_1} \psi_{i_2 j_2 k_2}&=&\psi_{i_2 j_2 k_2}\psi_{i_1 j_1 k_1}\cr
&=&-\psi_{i_1 j_1 k_1} \psi_{i_2 j_2 k_2}\cr
&=&{\rm sgn}\left(\,  (12)\,\right)\psi_{i_1 j_1 k_1} \psi_{i_2 j_2 k_2}
\eea
since the fermions are described using Grassman numbers.
The equations (\ref{bos}),(\ref{fer}) will be important in the next section.

\subsection{Counting and construction for bosonic tensors}

Our goal in this section is to count the number of gauge invariant operators that can be constructed from the 
bosonic tensors introduced above.
To construct gauge invariants we need to completely contract the indices of $\Phi_{IJK}$ with the indices of
$\bar\Phi^{IJK}$.
In general, this is accomplished using three permutations $\sigma_1,\sigma_2,\sigma_3\in S_n$ 
(or equivalently, one permutation $\sigma_1\circ\sigma_2\circ\sigma_3\in S_n\times S_n\times S_n$) as follows
\bea
\bar\Phi\,\cdot\,\sigma_1\circ\sigma_2\circ\sigma_3\,\cdot\,\Phi
=\bar\Phi^{IJK}\Phi_{\sigma_1(I)\sigma_2(J)\sigma_3(K)}\label{invariants}
\eea
%
The invariants given in equation (\ref{invariants}) are over complete: the $\phi$'s and $\bar\phi$'s are bosons, so 
we have the symmetry given in (\ref{bos}) which must be accounted for.
Let $\beta_1\in S_n$ be an arbitrary permutation of the $\phi$'s and let $\beta_2\in S_n$ be an arbitrary permutation 
of the $\bar\phi$'s. 
Then (we act to the right if we act on lower indices and to the left if we act on upper indices)
\bea
\bar\Phi\,\cdot\,\sigma_1\circ\sigma_2\circ\sigma_3\,\cdot\,\Phi=
(\bar\Phi\cdot\beta_2)\,\cdot\,\sigma_1\circ\sigma_2\circ\sigma_3\,\cdot\,(\beta_1\cdot\Phi)
\eea
Manipulating this a little, we have\footnote{A useful identity to keep in mind is the following:
$\bar\Phi^{\gamma^{-1}(K)}\Phi_K =\bar\Phi^K \Phi_{\gamma (K)}$.
This follows very simply by using the explicit representation 
$(\sigma)^I_J=\delta^{i_1}_{j_{\sigma (1)}}\cdots \delta^{i_n}_{j_{\sigma (n)}}$.}
\bea
\Phi_{\sigma_1(I)\sigma_2(J)\sigma_3(K)}\bar\Phi^{IJK}&=&
\Phi_{\sigma_1(\beta_1(I))\sigma_2(\beta_1(J))\sigma_3(\beta_1(K))}\bar\Phi^{\beta_2(I)\beta_2(J)\beta_2(K)}\cr
&=&\Phi_{\beta_2^{-1}(\sigma_1(\beta_1(I)))\beta_2^{-1}(\sigma_2(\beta_1(J)))\beta_2^{-1}(\sigma_3(\beta_1(K)))}
\bar\Phi^{IJK}
\eea
Thus, $(\sigma_1,\sigma_2,\sigma_3)$ and 
$(\beta_1\sigma_1\beta_2,\beta_1\sigma_2\beta_2,\beta_1\sigma_3\beta_2)$
define the same gauge invariant operator.
This implies that we have one gauge invariant operator for each element in the double coset
\bea
 S_n\setminus S_n\times S_n\times S_n\, /\, S_n
\eea
This understanding of the structure of the space of gauge invariant observables was first achieved 
in \cite{Geloun:2013kta}.
The generalization to other ranks is obvious.
For example, rank 5 tensors would be elements of the coset
\bea
   S_n\setminus S_n\times S_n\times S_n\times S_n\times S_n\, /\, S_n
\eea
The number of elements in a double coset $|H_1\setminus G/H_2|$ is given, by Burnside's Lemma, as
\bea
   |H_1\setminus G/H_2| ={1\over |H_1||H_2|}\sum_{h_1\in H_1}\sum_{h_2\in H_2}\sum_{g\in G}
\delta(h_1 g h_2 g^{-1})
\eea
Thus, for example, the number ${\cal N}_3$ of rank 3 tensors built using $n$ fields is given by
\bea
{\cal N}_3 ={1\over (n!)^2}\sum_{\sigma_1,\sigma_2,\sigma_3\in S_n}
\sum_{\beta_1,\beta_2\in S_n}\delta (\beta_1\sigma_1 \beta_2\sigma_1^{-1})
\delta (\beta_1\sigma_2 \beta_2\sigma_2^{-1})\delta (\beta_1\sigma_3 \beta_2\sigma_3^{-1})
\eea
To make sure the generalization is clear, we simply quote the count for the number of rank $q$ tensors built
using $n$ fields
\bea
{\cal N}_q={1\over (n!)^2}\,\,\sum_{\sigma_1,\cdots,\sigma_q\in S_n}\,\,
\sum_{\beta_1,\beta_2\in S_n}\,\,\prod_{i=1}^q \delta (\beta_1\sigma_i \beta_2\sigma_i^{-1})
\eea

The arguments we have just outlined are not the most natural when we generalize to fermionic tensors. 
To perform the counting in a way that will generalize nicely to the fermionic case, we will change basis. The operators
\bea
{\cal O}(\sigma_1,\sigma_2,\sigma_3)=\bar\Phi\,\cdot\,\sigma_1\circ\sigma_2\circ\sigma_3\,\cdot\,\Phi
\eea
define the ``permutation basis''. We will Fourier transform to the representation theory basis as follows
\bea
&&({\cal O}_{r_1,r_2,r_3})_{\alpha_1\alpha_2\alpha_3,\beta_1\beta_2\beta_3}=\cr\cr
&&\qquad\sum_{\sigma_1,\sigma_2,\sigma_2}{\cal O}(\sigma_1,\sigma_2,\sigma_3)
\Gamma^{r_1}{}_{\alpha_1\beta_1}(\sigma_1)
\Gamma^{r_2}{}_{\alpha_2\beta_2}(\sigma_2)
\Gamma^{r_3}{}_{\alpha_3\beta_3}(\sigma_3)
\eea
All of the representations above are irreducible representations of $S_n$, i.e. $r_i\vdash n$, $i=1,2,3$.
We again have to deal with the symmetry present as a consequence of (\ref{bos}).
The simplest way to do this is to couple the row indices to the trivial irreducible representation and to couple the 
column indices to the trivial irreducible representation of the diagonal $S_n$.
The tensor product of the irreducible representations involved is
\bea
  V_{r_1}\otimes V_{r_2}\otimes V_{r_3}=\bigoplus_r g_{r_1\, r_2\, r_3\, r}V_r
\eea
The Kronecker coefficients $g_{r_1\, r_2\, r_3\, r}$ are non-negative integers that count how many times irreducible
representation $r$ appears in the tensor product $r_1\otimes r_2\otimes r_3$. 
To perform the projection to the trivial, introduce the branching coefficients 
$B^\gamma_{\alpha_1\alpha_2\alpha_3}$ defined by
\bea
   {1\over n!}\sum_{\sigma\in S_n}
\Gamma^{r_1}{}_{\alpha_1\beta_1}(\sigma)
\Gamma^{r_2}{}_{\alpha_2\beta_2}(\sigma)
\Gamma^{r_3}{}_{\alpha_3\beta_3}(\sigma)
=\sum_\gamma B^\gamma_{\alpha_1\alpha_2\alpha_3}B^\gamma_{\beta_1\beta_2\beta_3}
\eea
The branching coefficients provide an orthonormal basis for the subspace of $r_1\otimes r_2\otimes r_3$ that
carries the trivial representation, i.e.
\bea
B^{\gamma_1}_{\alpha_1\alpha_2\alpha_3}B^{\gamma_2}_{\alpha_1\alpha_2\alpha_3}
=\delta^{\gamma_1\gamma_2}
\eea
and where we employ the usual convention that repeated indices are summed.
The gauge invariant operators are now given by
\bea
{\cal O}_{r_1,r_2,r_3}^{\gamma_1\gamma_2}=
B^{\gamma_1}_{\alpha_1\alpha_2\alpha_3}
({\cal O}_{r_1,r_2,r_3})_{\alpha_1\alpha_2\alpha_3,\beta_1\beta_2\beta_3}
B^{\gamma_2}_{\beta_1\beta_2\beta_3}
\eea
We will also write this as
\bea
{\cal O}_{r_1,r_2,r_3}^{\gamma_1\gamma_2}=
\sum_{\sigma_1\in S_n}\sum_{\sigma_2\in S_n}\sum_{\sigma_3\in S_n}
C_{r_1,r_2,r_3}^{\gamma_1\gamma_2}(\sigma_1,\sigma_2,\sigma_3)
{\cal O}(\sigma_1,\sigma_2,\sigma_3)\label{btrsp}
\eea
where
\bea
C_{r_1,r_2,r_3}^{\gamma_1\gamma_2}(\sigma_1,\sigma_2,\sigma_3)=
B^{\gamma_1}_{\alpha_1\alpha_2\alpha_3}
\Gamma^{r_1}{}_{\alpha_1\beta_1}(\sigma_1)
\Gamma^{r_2}{}_{\alpha_2\beta_2}(\sigma_2)
\Gamma^{r_3}{}_{\alpha_3\beta_3}(\sigma_3)
B^{\gamma_2}_{\beta_1\beta_2\beta_3}
\eea
is in fact a restricted character, in the language introduced in \cite{de Mello Koch:2007uu},\cite{Pasukonis:2013ts}.
Thus, (\ref{btrsp}) provides the restricted Schur polynomial basis for the gauge invariant operators of the bosonic
tensor model.

Since each multiplicity runs from $1$ to $g_{r_1\, r_2\, r_3\, 1}$ and each operator is labeled by a pair of
multiplicity labels, this second construction shows that the number of gauge invariant operators, constructed
using $n$ $\phi$'s and $n$ $\bar\phi$'s, is given by
\bea
\sum_{r_i\vdash n\,\, l(r_i)\le N_i}g_{r_1\, r_2\, r_3\, 1}^2\label{cftbc}
\eea
where we have used $1$ to denote the rep labeled by a Young diagram with a single row of $n$ boxes.
This is in complete agreement with \cite{will}, as already pointed out in \cite{Diaz:2017kub}.
A standard result which follows from the orthogonality of characters is
\bea
g_{r_1\, r_2\, r_3\, 1}={1\over n!}\sum_{\sigma\in S_n}
\chi_{r_1}(\sigma)\chi_{r_2}(\sigma)\chi_{r_3}(\sigma)
=g_{r_1\, r_2\, r_3}
\eea
Some checks of the counting formula (\ref{cftbc}) are given in Appendix \ref{CheckCount}.

\subsection{Counting and construction for fermionic tensors}

Our goal in this section is to count the number of gauge invariant operators that can be constructed from the 
fermionic tensors introduced above.
To construct gauge invariants we need to completely contract the indices of $\Psi_{IJK}$ with the indices of
$\bar\Psi^{IJK}$.
In general, this is again accomplished using three permutations $\sigma_1,\sigma_2,\sigma_3\in S_n$ 
(or equivalently $\sigma_1\circ\sigma_2\circ\sigma_3\in S_n\times S_n\times S_n$) as follows
\bea
\bar\Psi\,\cdot\,\sigma_1\circ\sigma_2\circ\sigma_3\,\cdot\,\Psi
=\bar\Psi^{IJK}\Psi_{\sigma_1(I)\sigma_2(J)\sigma_3(K)}\label{finvariants}
\eea
The invariants given in equation (\ref{finvariants}) are again over complete: the $\psi$'s and $\bar\psi$'s are fermions,
so we have the symmetry given in (\ref{fer}) which must be accounted for.
Following our discussion for the bosons, let $\beta_1\in S_n$ be an arbitrary permutation of the $\psi$'s and let
$\beta_2\in S_n$ be an arbitrary permutation of the $\bar\psi$'s. 
Then (exactly as for bosonic tensors, we act to the right if we act on lower indices and to the left, if we act on upper
indices)
\bea
\bar\Psi\,\cdot\,\sigma_1\circ\sigma_2\circ\sigma_3\,\cdot\,\Psi
={\rm sgn}(\beta_1){\rm sgn}(\beta_2)
(\bar\Psi\cdot\beta_2)\,\cdot\,\sigma_1\circ\sigma_2\circ\sigma_3\,\cdot\,(\beta_1\cdot\Psi)
\eea
Manipulating this a little, we have
\bea
\Psi_{\sigma_1(I)\sigma_2(J)\sigma_3(K)}\bar\Psi^{IJK}&=&
{\rm sgn}(\beta_1){\rm sgn}(\beta_2)
\Psi_{\sigma_1(\beta_1(I))\sigma_2(\beta_1(J))\sigma_3(\beta_1(K))}
\bar\Psi^{\beta_2(I)\beta_2(J)\beta_2(K)}\cr
&=&{\rm sgn}(\beta_1){\rm sgn}(\beta_2)
\Psi_{\beta_2^{-1}(\sigma_1(\beta_1(I)))\beta_2^{-1}(\sigma_2(\beta_1(J)))\beta_2^{-1}(\sigma_3(\beta_1(K)))}
\bar\Psi^{IJK}\cr
&&
\eea
We will still have to account for this symmetry.
To do this, it again proves useful to change basis. The operators
\bea
{\cal P}(\sigma_1,\sigma_2,\sigma_3)=\bar\Psi\,\cdot\,\sigma_1\circ\sigma_2\circ\sigma_3\,\cdot\,\Psi
\eea
define the ``permutation basis''. Again, Fourier transform to the representation theory basis as follows
\bea
&&({\cal P}_{r_1,r_2,r_3})_{\alpha_1\alpha_2\alpha_3,\beta_1\beta_2\beta_3}=\cr\cr
&&\qquad\sum_{\sigma_1,\sigma_2,\sigma_2}{\cal P}(\sigma_1,\sigma_2,\sigma_3)
\Gamma^{r_1}{}_{\alpha_1\beta_1}(\sigma_1)
\Gamma^{r_2}{}_{\alpha_2\beta_2}(\sigma_2)
\Gamma^{r_3}{}_{\alpha_3\beta_3}(\sigma_3)
\eea
We now have to deal with the symmetry present as a consequence of (\ref{fer}).
The simplest way to do this is to couple the row indices to the antisymmetric irreducible representation 
and to couple the column indices to the antisymmetric irreducible representation of the diagonal $S_n$.
By the antisymmetric irreducible representation, (denoted $(1^n)$) we mean the irreducible representation 
labeled by a Young diagram that has a single column of $n$ boxes.
This is a one dimensional representation defined by
\bea
\Gamma^{(1^n)}(\sigma)={\rm sgn}(\sigma)
\eea
To perform the projection to the antisymmetric irreducible representation, we again introduce branching coefficients
\bea
   {1\over n!}\sum_{\sigma\in S_n}
\Gamma^{r_1}{}_{\alpha_1\beta_1}(\sigma)
\Gamma^{r_2}{}_{\alpha_2\beta_2}(\sigma)
\Gamma^{r_3}{}_{\alpha_3\beta_3}(\sigma)
{\rm sgn}(\sigma)
=\sum_\gamma \tilde B^\gamma_{\alpha_1\alpha_2\alpha_3}\tilde B^\gamma_{\beta_1\beta_2\beta_3}
\eea
We are using a tilde to distinguish the branching coefficients defined using the antisymmetric irreducible representation,
from those relevant for the bosons which are defined using the symmetric representation. 
The branching coefficients again define an orthonormal basis
\bea
\tilde B^{\gamma_1}_{\alpha_1\alpha_2\alpha_3}\tilde B^{\gamma_2}_{\alpha_1\alpha_2\alpha_3}
=\delta^{\gamma_1\gamma_2}
\eea
The gauge invariant operators are now given by
\bea
{\cal P}_{r_1,r_2,r_3}^{\gamma_1\gamma_2}=
\tilde B^{\gamma_1}_{\alpha_1\alpha_2\alpha_3}
({\cal P}_{r_1,r_2,r_3})_{\alpha_1\alpha_2\alpha_3,\beta_1\beta_2\beta_3}
\tilde B^{\gamma_2}_{\beta_1\beta_2\beta_3}
\eea
Once again, we can write this as
\bea
{\cal P}_{r_1,r_2,r_3}^{\gamma_1\gamma_2}=
\sum_{\sigma_1\in S_n}\sum_{\sigma_2\in S_n}\sum_{\sigma_3\in S_n}
\tilde C_{r_1,r_2,r_3}^{\gamma_1\gamma_2}(\sigma_1,\sigma_2,\sigma_3)
{\cal P}(\sigma_1,\sigma_2,\sigma_3)\label{ftrsp}
\eea
where
\bea
\tilde C_{r_1,r_2,r_3}^{\gamma_1\gamma_2}(\sigma_1,\sigma_2,\sigma_3)=
\tilde B^{\gamma_1}_{\alpha_1\alpha_2\alpha_3}
\Gamma^{r_1}{}_{\alpha_1\beta_1}(\sigma_1)
\Gamma^{r_2}{}_{\alpha_2\beta_2}(\sigma_2)
\Gamma^{r_3}{}_{\alpha_3\beta_3}(\sigma_3)
\tilde B^{\gamma_2}_{\beta_1\beta_2\beta_3}
\eea
is again a restricted character.
Thus, (\ref{ftrsp}) provides the restricted Schur polynomial basis for the gauge invariant operators of the 
fermionic tensor model.

This construction shows that the number of gauge invariant operators is given by
\bea
\sum_{r_i\vdash n\,\, l(r_i)\le N_i}g_{r_1\, r_2\, r_3\, (1^n)}^2
\eea
A standard result which follows from the orthogonality of characters is
\bea
g_{r_1\, r_2\, r_3\, (1^n)}={1\over n!}\sum_{\sigma\in S_n}
\chi_{r_1}(\sigma)\chi_{r_2}(\sigma)\chi_{r_3}(\sigma){\rm sgn}(\sigma)
\eea
Some checks of this counting formula are given in Appendix \ref{CheckCount}.

\section{Correlators of Gauge Invariant Operators}

In this section we will compute the correlation functions of the operators defined in the previous section.
Since these operators are neutral under the gauge symmetry they can develop a nonzero vacuum expectation
value.
It is interesting to compute these values as their large $N$ limit must be reproduced by the classical equations 
of motion of collective field theory.
We also compute the two point functions of normal ordered gauge invariant operators.
The large $N$ limit of these two point functions must be reproduced by considering quadratic fluctuations
about the classical collective configuration.

\subsection{Bosonic Correlators}

The free field two point function is
\bea
\langle\bar\phi^{ijk}\phi_{lmn}\rangle=\delta^i_l\delta^j_m\delta^k_n\label{ettpf}
\eea
This is valid both as a formula in a zero dimensional random tensor model, or as an equal time
two point function in the tensor model quantum mechanics. 
Wick's theorem can be written as
\bea
\langle\bar\Phi^{IJK}\Phi_{LMN}\rangle &=&\sum_{\sigma\in S_n}\prod_{a=1}^n
\delta^{i_a}_{l_{\sigma(a)}}\delta^{j_a}_{m_{\sigma(a)}}\delta^{k_a}_{n_{\sigma(a)}}\cr
&=&\sum_{\sigma\in S_n}(\sigma)^I_L(\sigma)^J_M(\sigma)^K_N
\eea
There are two interesting correlators to consider: first we could consider the one point functions
$\langle {\cal O}^{\gamma_1\gamma_2}_{r_1,r_2,r_3}\rangle$; second we could consider the two
point function of normal ordered operators $\langle :{\cal O}^{\gamma_1\gamma_2}_{r_1 r_2 r_3}:\,
:{\cal O}^{\gamma_3\gamma_4}_{s_1 s_2 s_3}: \rangle$.

{\vskip 0.3cm}

\noindent
{\bf One point functions:}
We will use the fact that 
\bea
   {\rm Tr}_{V_j}(\sigma)=\delta^{i_1}_{i_{\sigma (1)}}\cdots \delta^{i_n}_{i_{\sigma (n)}}=N^{C(\sigma)}_j
\eea
where $C(\sigma)$ denotes the number of cycles in the permutation $\sigma$.
In addition, we will use the orthogonality relation
\bea
{d_r\over n!}\sum_{\sigma\in S_n}\Gamma_r(\sigma)_{ab}\Gamma_s(\sigma^{-1})_{cd}
=\delta_{rs}\delta_{bc}\delta_{ad}\label{or}
\eea
to obtain
\bea
{d_r\over n!}\sum_{s\vdash n}\sum_{\sigma_1\in S_n}\Gamma_r(\sigma_1 )_{cd}
\chi_s (\sigma\sigma_1^{-1})=\Gamma_{r}(\sigma)_{cd}
\eea
Finally, we will use the relation, valid for Schur polynomials
\bea
   {\rm Tr}(\sigma Z)=\sum_R\chi_R(\sigma)\chi_R(Z)
\eea
evaluated at $Z=1$ to find
\bea
   {\rm Tr}(\sigma)={1\over n!}\sum_{R\vdash n} d_R\chi_R(\sigma)f_R(N)
\eea
$f_R(N)$ is the product of the factors of Young diagram $R$ understood as a representation of $U(N)$.
Recall that the factor of a box in row $i$ and column $j$ is $N-i+j$.
We use $\chi_R(\sigma)$ to denote a character of the symmetric group and $\chi_R(Z)$ to denote
a Schur polynomial. 
The two are distinguished only by their argument, which is either an element of the
symmetric group $\sigma\in S_n$ or an $N\times N$ matrix $Z$.
We are now ready to compute the one point function
\bea
\langle {\cal O}^{\gamma_1\gamma_2}_{r_1 r_2 r_3}\rangle
&=&\sum_{\sigma_i\in S_n}\langle\bar\Phi\cdot \sigma_1\circ\sigma_2\circ\sigma_3\cdot\Phi\rangle
B^{\gamma_1} \Gamma_{r_1}(\sigma_1)\Gamma_{r_2}(\sigma_2)\Gamma_{r_3}(\sigma_3)B^{\gamma_2}\cr\cr
&=&\sum_{\sigma ,\sigma_i\in S_n}N_1^{C(\sigma\sigma_1)}
N_2^{C(\sigma\sigma_2)}
N_3^{C(\sigma\sigma_3)}
B^{\gamma_1} \Gamma_{r_1}(\sigma_1)\Gamma_{r_2}(\sigma_2)\Gamma_{r_3}(\sigma_3)B^{\gamma_2}\cr\cr
&=&\sum_{\sigma ,\sigma_i\in S_n}\sum_{s_i\vdash n}\left({1\over n!}\right)^3
d_{s_1}\chi_{s_1}(\sigma\sigma_1^{-1})f_{s_1}(N_1)
d_{s_2}\chi_{s_2}(\sigma\sigma_2^{-1})f_{s_2}(N_2)\cr\cr
&&d_{s_3}\chi_{s_3}(\sigma\sigma_3^{-1})f_{s_3}(N_3)
B^{\gamma_1} \Gamma_{r_1}(\sigma_1)\Gamma_{r_2}(\sigma_2)\Gamma_{r_3}(\sigma_3)B^{\gamma_2}\cr\cr
&=&\sum_{\sigma\in S_n}f_{r_1}(N_1)f_{r_2}(N_2)f_{r_3}(N_3)
B^{\gamma_1} \Gamma_{r_1}(\sigma)\Gamma_{r_2}(\sigma)\Gamma_{r_3}(\sigma)B^{\gamma_2}\cr\cr
&=&n! f_{r_1}(N_1)f_{r_2}(N_2)f_{r_3}(N_3)\delta^{\gamma_1\gamma_2}\label{bos1pnt}
\eea
See Appendix \ref{Ops} for some checks of this formula.

{\vskip 0.3cm}

\noindent
{\bf Two point functions of normal ordered operators:} Using the identities given above, it is straightforward to compute
\bea
&&\langle :{\cal O}^{\gamma_1\gamma_2}_{r_1 r_2 r_3}:\,
:{\cal O}^{\gamma_3\gamma_4}_{s_1 s_2 s_3}: \rangle
=\sum_{\sigma\in S_n}
\sum_{\rho \in S_n}
\sum_{\sigma_i\in S_n}
\sum_{\tau_i\in S_n}
{\rm Tr}(\sigma_1\sigma\tau_1\rho)
{\rm Tr}(\sigma_2\sigma\tau_2\rho)
{\rm Tr}(\sigma_3\sigma\tau_3\rho)\cr\cr
&&\times B^{\gamma_1}\Gamma_{r_1}(\sigma_1)\Gamma_{r_2}(\sigma_2)\Gamma_{r_3}(\sigma_3)B^{\gamma_2}
\,\, B^{\gamma_3}\Gamma_{s_1}(\tau_1)\Gamma_{s_2}(\tau_2)\Gamma_{s_3}(\tau_3)B^{\gamma_4}
\cr\cr
&&=\sum_{\sigma \rho \sigma_i \tau_i}
N^{C(\sigma_1\sigma\tau_1\rho)}
N^{C(\sigma_2\sigma\tau_2\rho)}
N^{(\sigma_3\sigma\tau_3\rho)}\cr\cr
&&\times B^{\gamma_1}\Gamma_{r_1}(\sigma_1)\Gamma_{r_2}(\sigma_2)\Gamma_{r_3}(\sigma_3)B^{\gamma_2}
\,\, B^{\gamma_3}\Gamma_{s_1}(\tau_1)\Gamma_{s_2}(\tau_2)\Gamma_{s_3}(\tau_3)B^{\gamma_4}
\cr\cr
&&=\sum_{\sigma \rho \sigma_i \tau_i}\sum_{t_i\vdash n}\left({1\over n!}\right)^3
d_{t_1}\chi_{t_1}(\sigma_1\sigma\tau_1\rho)f_{t_1}(N_1)
d_{t_2}\chi_{t_2}(\sigma_2\sigma\tau_2\rho)f_{t_2}(N_2)
d_{t_3}\chi_{t_3}(\sigma_3\sigma\tau_3\rho)f_{t_3}(N_3)\cr\cr
&&\times B^{\gamma_1}\Gamma_{r_1}(\sigma_1)\Gamma_{r_2}(\sigma_2)\Gamma_{r_3}(\sigma_3)B^{\gamma_2}
\,\, B^{\gamma_3}\Gamma_{s_1}(\tau_1)\Gamma_{s_2}(\tau_2)\Gamma_{s_3}(\tau_3)B^{\gamma_4}
\cr\cr
&&=(n!)^2\delta_{r_1s_1}\delta_{r_2s_2}\delta_{r_3s_3}f_{r_1}(N_1)f_{r_2}(N_2)f_{r_3}(N_3)
{n!\over d_{r_1}}{n!\over d_{r_2}}{n!\over d_{r_3}}\delta^{\gamma_1\gamma_4}\delta^{\gamma_2\gamma_3}
\label{bos2pnt}
\eea
Some checks of this formula are given in Appendix \ref{Ops}.

\subsection{Fermionic Correlators}

The relevant two point function for the fermionic tensor model is
\bea
 \langle \bar\psi^{ijk}\psi_{lmn} \rangle = \delta^i_l\delta^j_m\delta^k_n
\eea
This is valid, as for the bosons, both as a formula in a zero dimensional random tensor model, or as an equal time
two point function in the tensor model quantum mechanics. 
Since the fermionic fields anticommute, it is important to spell out the ordering of the fields.
Order the fields in the following way
\bea
\bar\Psi^{IJK}
  \Psi_{LMN}  = \bar \psi^{i_1 j_1 k_1}\bar\psi^{i_2 j_2 k_2}\cdots \bar\psi^{i_n j_n k_n}
   \psi_{l_n m_n n_n}\cdots \psi_{l_2 m_2 n_2} \psi_{l_1 m_1 n_1}
\eea
With this ordering spelled out, a simple application of Wick's theorem now gives
\bea
  \langle \bar\psi^{IJK} \psi_{LMN}\rangle 
=\sum_{\sigma\in S_n} {\rm sgn}(\sigma)\sigma^I_L \sigma^J_M \sigma^K_N
\eea

{\vskip 0.3cm}

\noindent
{\bf One point functions:} A simple computation shows that
\bea
\langle {\cal P}^{\gamma_1\gamma_2}_{r_1 r_2 r_3}\rangle
&=&\sum_{\sigma_i\in S_n}\langle\bar\Psi\cdot \sigma_1\circ\sigma_2\circ\sigma_3\cdot\Psi\rangle
\tilde B^{\gamma_1} \Gamma_{r_1}(\sigma_1)\Gamma_{r_2}(\sigma_2)\Gamma_{r_3}(\sigma_3)\tilde B^{\gamma_2}\cr\cr
&=&\sum_{\sigma ,\sigma_i\in S_n}{\rm sgn}(\sigma)
N_1^{C(\sigma\sigma_1)}
N_2^{C(\sigma\sigma_2)}
N_3^{C(\sigma\sigma_3)}
\tilde B^{\gamma_1} \Gamma_{r_1}(\sigma_1)\Gamma_{r_2}(\sigma_2)\Gamma_{r_3}(\sigma_3)\tilde B^{\gamma_2}\cr\cr
&=&\sum_{\sigma\in S_n}{\rm sgn}(\sigma)f_{r_1}(N_1)f_{r_2}(N_2)f_{r_3}(N_3)
\tilde B^{\gamma_1} \Gamma_{r_1}(\sigma)\Gamma_{r_2}(\sigma)\Gamma_{r_3}(\sigma)\tilde B^{\gamma_2}\cr\cr
&=&n! f_{r_1}(N_1)f_{r_2}(N_2)f_{r_3}(N_3)\delta^{\gamma_1\gamma_2}\label{fermion1pt}
\eea
See Appendix \ref{Ops} for examples and checks of this formula.

{\vskip 0.3cm}

\noindent
{\bf Two point functions of normal ordered operators:} Using the identities given above
\bea
&&\langle :{\cal P}^{\gamma_1\gamma_2}_{r_1 r_2 r_3}:\,
:{\cal P}^{\gamma_3\gamma_4}_{s_1 s_2 s_3}: \rangle
=\sum_{\sigma\in S_n}
\sum_{\rho \in S_n}
\sum_{\sigma_i\in S_n}
\sum_{\tau_i\in S_n}{\rm sgn}(\sigma){\rm sgn}(\rho)
{\rm Tr}(\sigma_1\sigma\tau_1\rho)
{\rm Tr}(\sigma_2\sigma\tau_2\rho)
{\rm Tr}(\sigma_3\sigma\tau_3\rho)\cr\cr
&&\times \tilde B^{\gamma_1}\Gamma_{r_1}(\sigma_1)\Gamma_{r_2}(\sigma_2)\Gamma_{r_3}(\sigma_3)\tilde B^{\gamma_2}
\,\, \tilde B^{\gamma_3}\Gamma_{s_1}(\tau_1)\Gamma_{s_2}(\tau_2)\Gamma_{s_3}(\tau_3)\tilde B^{\gamma_4}
\cr\cr
&&=\sum_{\sigma \rho \sigma_i \tau_i}{\rm sgn}(\sigma){\rm sgn}(\rho)
N^{C(\sigma_1\sigma\tau_1\rho)}
N^{C(\sigma_2\sigma\tau_2\rho)}
N^{(\sigma_3\sigma\tau_3\rho)}\cr\cr
&&\times \tilde B^{\gamma_1}\Gamma_{r_1}(\sigma_1)\Gamma_{r_2}(\sigma_2)\Gamma_{r_3}(\sigma_3)\tilde B^{\gamma_2}
\,\, \tilde B^{\gamma_3}\Gamma_{s_1}(\tau_1)\Gamma_{s_2}(\tau_2)\Gamma_{s_3}(\tau_3)\tilde B^{\gamma_4}
\cr\cr
&&=(n!)^2\delta_{r_1s_1}\delta_{r_2s_2}\delta_{r_3s_3}f_{r_1}(N_1)f_{r_2}(N_2)f_{r_3}(N_3)
{n!\over d_{r_1}}{n!\over d_{r_2}}{n!\over d_{r_3}}
\delta^{\gamma_1\gamma_4}\delta^{\gamma_2\gamma_3}
\label{fermion2pt}
\eea
Appendix \ref{Ops} illustrates and checks this formula in some simple cases.

\section{Algebra of the Gauge Invariant Operators}

The gauge invariant operators that we have introduced above close an interesting algebra: we will argue that
the gauge invariant operators have a ring structure.
Algebras of gauge invariant operators have also been considered in \cite{Mattioli:2016eyp}.
To develop the algebra for our tensor model, we will need to develop some properties of the restricted character.
We can always assume that we work in an orthogonal representation of the symmetric group.
In this case the restricted characters obey
\bea
C^{\gamma_1\gamma_2}_{r_1r_2r_3}(\sigma_1,\sigma_2,\sigma_3)
=C^{\gamma_2\gamma_1}_{r_1r_2r_3}(\sigma_1^{-1},\sigma_2^{-1},\sigma_3^{-1})\cr\cr
\tilde C^{\gamma_1\gamma_2}_{r_1r_2r_3}(\sigma_1,\sigma_2,\sigma_3)
=\tilde C^{\gamma_2\gamma_1}_{r_1r_2r_3}(\sigma_1^{-1},\sigma_2^{-1},\sigma_3^{-1})
\eea
They also enjoy a ``completeness identity'' given by
\bea 
\sum_{\sigma_1\in S_n}\sum_{\sigma_2\in S_n}\sum_{\sigma_3\in S_n}
C^{\gamma_1\gamma_2}_{r_1r_2r_3}(\sigma_1,\sigma_2,\sigma_3)
C^{\gamma_3\gamma_4}_{s_1s_2s_3}(\sigma_1,\sigma_2,\sigma_3)
={n!\over d_{r_1}}{n!\over d_{r_2}}{n!\over d_{r_3}}
\delta_{r_1 s_1}\delta_{r_2 s_2}\delta_{r_3 s_3}\delta^{\gamma_1 \gamma_3}
\delta^{\gamma_2\gamma_4}\cr\cr
\sum_{\sigma_1\in S_n}\sum_{\sigma_2\in S_n}\sum_{\sigma_3\in S_n}
\tilde C^{\gamma_1\gamma_2}_{r_1r_2r_3}(\sigma_1,\sigma_2,\sigma_3)
\tilde C^{\gamma_3\gamma_4}_{s_1s_2s_3}(\sigma_1,\sigma_2,\sigma_3)
={n!\over d_{r_1}}{n!\over d_{r_2}}{n!\over d_{r_3}}
\delta_{r_1 s_1}\delta_{r_2 s_2}\delta_{r_3 s_3}\delta^{\gamma_1 \gamma_3}
\delta^{\gamma_2\gamma_4}
\eea
Using these formulas, we find the following interesting Fourier transform pairs
\bea
{\cal O}^{\gamma_1\gamma_2}_{r_1r_2r_3}=
\sum_{\sigma_1\in S_n}\sum_{\sigma_2\in S_n}\sum_{\sigma_3\in S_n}
C^{\gamma_1\gamma_2}_{r_1r_2r_3}(\sigma_1,\sigma_2,\sigma_3)
{\cal O}(\sigma_1,\sigma_2,\sigma_3)
\eea
\bea
{\cal O}(\sigma_1,\sigma_2,\sigma_3)=
\sum_{s_1\vdash n}\sum_{s_2\vdash n}\sum_{s_3\vdash n}\sum_{\gamma_1,\gamma_2}
{d_{s_1}\over n!}{d_{s_2}\over n!}{d_{s_3}\over n!}
C^{\gamma_1\gamma_2}_{s_1s_2s_3}(\sigma_1,\sigma_2,\sigma_3)
{\cal O}^{\gamma_1\gamma_2}_{s_1s_2s_3}
\eea
\bea
{\cal P}^{\gamma_1\gamma_2}_{r_1r_2r_3}=
\sum_{\sigma_1\in S_n}\sum_{\sigma_2\in S_n}\sum_{\sigma_3\in S_n}
\tilde C^{\gamma_1\gamma_2}_{r_1r_2r_3}(\sigma_1,\sigma_2,\sigma_3)
{\cal P}(\sigma_1,\sigma_2,\sigma_3)
\eea
\bea
{\cal P}(\sigma_1,\sigma_2,\sigma_3)
=\sum_{s_1\vdash n}\sum_{s_2\vdash n}\sum_{s_3\vdash n}\sum_{\gamma_1,\gamma_2}
{d_{s_1}\over n!}{d_{s_2}\over n!}{d_{s_3}\over n!}
\tilde C^{\gamma_1\gamma_2}_{s_1s_2s_3}(\sigma_1,\sigma_2,\sigma_3)
{\cal P}^{\gamma_1\gamma_2}_{s_1s_2s_3}
\eea
These formulas provide the clearest way to understand the relation between the permutation and representation
theory bases.

Now, in the permutation basis the gauge invariant operators close the following algebra
\bea
{\cal O}(\sigma_1,\sigma_2,\sigma_3){\cal O}(\rho_1,\rho_2,\rho_3)
={\cal O}(\sigma_1\circ\rho_1,\sigma_2\circ\rho_2,\sigma_3\circ\rho_3)
\eea
\bea
{\cal P}(\sigma_1,\sigma_2,\sigma_3){\cal P}(\rho_1,\rho_2,\rho_3)
={\cal P}(\sigma_1\circ\rho_1,\sigma_2\circ\rho_2,\sigma_3\circ\rho_3)
\eea
where $\sigma_i\in S_n$ and $\rho_i\in S_m$ for $i=1,2,3$.
Note that thanks to the way that we have ordered the fermions there are no $-1$ factors in this second
equation.

We can now work out the details of this algebra in the representation basis. 
A straightforward computation shows
\bea
{\cal O}^{\gamma_1\gamma_2}_{r_1 r_2 r_3}
{\cal O}^{\gamma_3\gamma_4}_{s_1 s_2 s_3}
=\sum_{t_1\vdash n+m}\sum_{t_2\vdash n+m}\sum_{t_3\vdash n+m}\sum_{\gamma_5\gamma_6}
f^{t_1 t_2 t_3;\gamma_1\gamma_2\gamma_3\gamma_4}_{r_1r_2r_3s_1s_2s_3;\gamma_5\gamma_6}
{\cal O}^{\gamma_5\gamma_6}_{t_1 t_2 t_3}
\eea
where $r_i\vdash n$ and $s_i\vdash m$ for $i=1,2,3$.
The structure constants for this algebra are given by
\bea
&&f^{t_1 t_2 t_3;\gamma_1\gamma_2\gamma_3\gamma_4}_{r_1r_2r_3s_1s_2s_3;\gamma_5\gamma_6} =
{d_{t_1}\over (n+m)!}{d_{t_2}\over (n+m)!}{d_{t_3}\over (n+m)!}
\sum_{\sigma_1\in S_n}\sum_{\sigma_2\in S_n}\sum_{\sigma_3\in S_n}
\sum_{\rho_1\in S_m}\sum_{\rho_2\in S_m}\sum_{\rho_3\in S_m}\cr\cr
&&\qquad C^{\gamma_1\gamma_2}_{r_1r_2r_3}(\sigma_1,\sigma_2,\sigma_3)
C^{\gamma_3\gamma_4}_{s_1s_2s_3}(\rho_1,\rho_2,\rho_3)
C^{\gamma_5\gamma_6}_{t_1t_2t_3}(\sigma_1\circ\rho_1,\sigma_2\circ\rho_2,\sigma_3\circ\rho_2)\cr\cr
&&\qquad\qquad=
{d_{t_1}n!m!\over (n+m)!d_{r_1}d_{s_1}}
{d_{t_2}n!m!\over (n+m)!d_{r_2}d_{s_2}}
{d_{t_3}n!m!\over (n+m)!d_{r_3}d_{s_3}}
B^{\gamma_1}_a\circ B^{\gamma_3}_b B^{\gamma_5}_{ab} \,\,\,
B^{\gamma_6}_{cd} B^{\gamma_2}_c\circ B^{\gamma_4}_d\cr
&&
\eea
To get to the last line above, we have simply performed the sum over the $\sigma_i$ and the $\rho_i$
using the orthogonality relation (\ref{or}). 
Remarkably, the structure constants are simply related to overlaps between branching coefficients!
Computing these overlaps is a well defined problem in the representation theory of the symmetric group.
Notice also that the structure constant, up to an overall factor, factorizes into a product of two overlaps of 
branching coefficients.

There is a similar algebra for the fermionic operators
\bea
{\cal P}^{\gamma_1\gamma_2}_{r_1 r_2 r_3}
{\cal P}^{\gamma_3\gamma_4}_{s_1 s_2 s_3}
=\sum_{t_1\vdash n+m}\sum_{t_2\vdash n+m}\sum_{t_3\vdash n+m}\sum_{\gamma_5\gamma_6}
g^{t_1 t_2 t_3;\gamma_1\gamma_2\gamma_3\gamma_4}_{r_1r_2r_3s_1s_2s_3;\gamma_5\gamma_6}
{\cal P}^{\gamma_5\gamma_6}_{t_1 t_2 t_3}
\eea
where the structure constants for this algebra are given by
\bea
&&g^{t_1 t_2 t_3;\gamma_1\gamma_2\gamma_3\gamma_4}_{r_1r_2r_3s_1s_2s_3;\gamma_5\gamma_6} =
{d_{t_1}\over (n+m)!}{d_{t_2}\over (n+m)!}{d_{t_3}\over (n+m)!}
\sum_{\sigma_1\in S_n}\sum_{\sigma_2\in S_n}\sum_{\sigma_3\in S_n}
\sum_{\rho_1\in S_m}\sum_{\rho_2\in S_m}\sum_{\rho_3\in S_m}\cr\cr
&&\qquad \tilde C^{\gamma_1\gamma_2}_{r_1r_2r_3}(\sigma_1,\sigma_2,\sigma_3)
\tilde C^{\gamma_3\gamma_4}_{s_1s_2s_3}(\rho_1,\rho_2,\rho_3)
\tilde C^{\gamma_5\gamma_6}_{t_1t_2t_3}(\sigma_1\circ\rho_1,\sigma_2\circ\rho_2,\sigma_3\circ\rho_2)\cr\cr
&&\qquad\qquad={d_{t_1}n!m!\over (n+m)!d_{r_1}d_{s_1}}
{d_{t_2}n!m!\over (n+m)!d_{r_2}d_{s_2}}
{d_{t_3}n!m!\over (n+m)!d_{r_3}d_{s_3}}
\tilde B^{\gamma_1}_a\circ \tilde B^{\gamma_3}_b \tilde B^{\gamma_5}_{ab} \,\,\,
\tilde B^{\gamma_6}_{cd} \tilde B^{\gamma_2}_c\circ \tilde B^{\gamma_4}_d\cr
&&
\eea
Again, the structure constants are simply related to overlaps between branching coefficients.

The existence of an algebraic structure for the gauge invariant operators has a remarkable consequence: we are
able to solve the free theory exactly.
To make this point, write the algebraic structure in a condensed notation as follows
\bea
{\cal O}^A{\cal O}^B=f^{AB}_C {\cal O}^C
\eea
with repeated indices summed.
At the risk of being pedantic, $A$ stands for two multiplicity labels ($\gamma_1,\gamma_2$ say) and three
Young diagrams ($r_1,r_2,r_3$ say).
Using this product repeatedly we find
\bea
\langle {\cal O}^{A_1}{\cal O}^{A_2}{\cal O}^{A_3}\cdots {\cal O}^{A_n}\rangle
=f^{A_1 A_2}_{C_1}f^{C_1 A_3}_{C_2}\cdots f^{C_{n-2} A_n}_{C_{n-1}}\langle {\cal O}^{C_{n-1}}\rangle
\eea
so that the computation of an $n$-point correlation function is reduced to the computation of a one point function.
We have already computed the most general one point function in the previous section.
Of course, the structure constants of the algebra need to be evaluated and this is non-trivial.
However, it does mean that the problem of solving the free tensor model has been reduced entirely to a problem
in $S_n$ representation theory.
This simplification is highly non-trivial.
There is a completely parallel argument for the fermionic tensor model.

\section{Collective Field Theory}

We have now constructed the complete set of gauge invariant variables and an algebra that these gauge invariants close.
In this section we would like to construct a (collective) field theory governing the dynamics of these variables.
Our discussion is guided by the dynamics of a single hermitian matrix $X=X^\dagger$ and we will review some
relevant background before we consider the collective field theory relevant for the tensor model.

A complete set of gauge invariant variables for the one matrix model is provided by the Schur polynomials\cite{Jevicki:1991yi}
\bea
   \chi_R (X)={1\over n!}\sum_{\sigma\in S_n}\chi_R(\sigma){\rm Tr} (\sigma X)
\eea
These variables again close an interesting algebra, given by
\bea
   \chi_R(X)\chi_S(X)=\sum_T g_{RST}\chi_T(X)
\eea
where $g_{RST}$ are the Littlewood-Richardson coefficients.
If we tried to quantize the Schur polynomial variables, it would be a mistake to treat them as independent, as the above
algebra proves.
In the case of a single matrix it is clear how we should proceed: one can select a smaller set of variables that are
independent
\bea
   \phi_n={\rm Tr}(X^n)\label{traces}
\eea
where we should restrict $n\le N$.
The complete set of gauge invariant variables, the Schur polynomials, are polynomials in the $\phi_n$.
This is an important point: the $\phi_n$ are the set of variables that are independent and by considering 
polynomials in these variables, we recover the complete set of gauge invariant operators.  
In the large $N$ limit, it is sensible to simply ignore the constraint $n\le N$\cite{collft}.
We can then consider the field
\bea
   \phi_k ={\rm Tr}(e^{ikX})
\eea
or its Fourier transform, $\phi (x)$.
The dynamics of this field, which is local in the emergent dimension $x$, is captured in the Das-Jevicki-Sakita
Hamiltonian\cite{collft,Das:1990kaa}.

We now want to explore the possibility that there is a similar description possible for tensor models.
Our first task is to identify the smaller set of independent variables which are independent and which we will quantize.
Further, by considering polynomials in these variables, we should reconstruct the complete set of gauge invariant
operators.

It proves convenient to work in the permutation basis
\bea
\bar\Phi\,\cdot\,\sigma_1\circ\sigma_2\circ\sigma_3\,\cdot\,\Phi
=\bar\Phi^{IJK}\Phi_{\sigma_1(I)\sigma_2(J)\sigma_3(K)}
\eea
These invariants were first counted by Geloun and Ramgoolam in \cite{Geloun:2013kta}.
They have identified the number of invariants with the series A110143 on the OEIS website.
This sequence counts the number of orbits obtained when $S_n$ acts on $S_n\times S_n$ via conjugacy, i.e.
for $g\in S_n$ and $(x,y)\in S_n\times S_n$ we have $g(x,y)=(gxg^{-1},gyg^{-1})$.
The number of invariants grows extremely rapidly
\bea
1, 4, 11, 43, 161, 901, 5579, 43206, 378360, 3742738,...
\eea
A useful way to label the invariants, following \cite{Geloun:2013kta}, is by bipartite cubic graphs with
edges labeled by the gauge group the corresponding index belongs to.

\begin{figure}[h]
        \centering
                \includegraphics[width=0.6\textwidth]{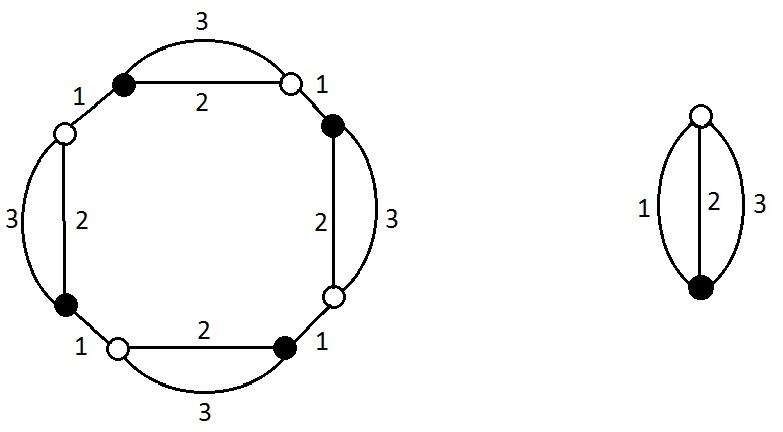}
        \label{dilpicjpg}
        \caption{The above figures label gauge invariant operators in the tensor model gauge theory. Black dots
                      correspond to $\bar\phi^{ijk}$'s and white dots to $\phi_{ijk}$s. A line labeled by $i$ is a gauge
                      index for $U(N_i)$. The operator on the left corresponds to
                      $\bar\phi^{i_1j_1k_1}\phi_{i_2j_1k_1}\bar\phi^{i_2j_2k_2}\phi_{i_3j_2k_2}\bar\phi^{i_3j_3k_3}\phi_{i_4j_3k_3}\bar\phi^{i_4j_4k_4}\phi_{i_1j_4k_4}$
and the operator on the right corresponds to $\bar\phi^{ijk}\phi_{ijk}$.}                  
\end{figure}

In the language of graphs it is easy to identify the smaller set of independent variables: they are the variables that
correspond to connected graphs.
The number of connected graphs can be counted using the plethystic logarithm and is identified with the series 
A057005 on the OEIS website\cite{Geloun:2013kta}.
The number of independent variables still grows extremely rapidly
\bea
1, 3, 7, 26, 97, 624, 4163, 34470, 314493, 3202839, 35704007, 433460014,...
\eea
This growth seems to be too rapid to manage.
We will now argue that we can restrict the dynamics to an even smaller set of variables.
To describe the smaller set of variables, it is useful to consider
\bea
  T^{i_1}{}_{i_2}=\bar\phi^{i_1jk}\phi_{i_2jk}
\eea
$T$ is a matrix on the vector space that carries the fundamental of $U(N_1)$.
It thus makes sense to take powers of $T$
\bea
   (T^n)^{i_1}{}_{i_2}=T^{i_1}{}_jT^j{}_k\cdots T^l{}_{i_2}
\eea
The smaller set of gauge invariants that we consider is given by
\bea
   \phi_n={\rm Tr}(T^n)
\eea
The Hamiltonian of the tensor model quantum mechanics we consider is given by
\bea
  H=-{\partial\over\partial\bar\phi^{ijk}}{\partial\over\partial\phi_{ijk}}+{1\over 4}\bar\phi^{ijk}\phi_{ijk}
\eea
The coefficient of the second term has been chosen to ensure that the equal time two point function is given 
by (\ref{ettpf}).
The kinetic terms of the Hamiltonian, when rewritten in terms of the new (collective) variables are
\bea
-{\partial\over\partial\bar\phi^{ijk}}{\partial\over\partial\phi_{ijk}}
=-\sum_{n,m}\Omega(n,m){\partial\over\partial\phi_n}{\partial\over\partial\phi_m}
+\sum_n\omega (n){\partial\over\partial\phi_n}
\eea
where\footnote{Note that these expressions are almost identical to the answers for the Hermitian one matrix
model which are $\Omega(n,m)=nm\phi_{n+m-2}$ and $\omega (n)=-\sum_{r=0}^{n-2}\phi_r\phi_{n-r-2}$.}
\bea
\Omega(n,m)={\partial\phi_n\over\partial\bar\phi^{ijk}}{\partial\phi_m\over\partial\phi_{ijk}}
=nm\phi_{n+m-1}
\eea
\bea
\omega (n)=-{\partial\over\partial\bar\phi^{ijk}}\left( {\partial\phi_n\over\partial\phi_{ijk}}\right)
=-\sum_{r=0}^{n-2}\phi_r\phi_{n-r-1}-N_2 N_3m\phi_{m-1}
\eea
It is nontrivial that $\Omega(n,m)$ and $\omega (n)$ can be expressed in terms of the $\phi_n$.
This implies that the Hamiltonian itself can be expressed in terms of this smaller set of variables, and hence that
it is consistent with the dynamics to restrict to this smaller set of variables.
When written in terms of the new variables, the Hamiltonian
\bea
H=-\sum_{n,m}\Omega(n,m){\partial\over\partial\phi_n}{\partial\over\partial\phi_m}
+\sum_n\omega (n){\partial\over\partial\phi_n}+{1\over 4}\phi_1
\eea
is not hermitian.
This simply reflects the fact that in the new variables the inner product is accompanied by a non-trivial Jacobian
$J[\phi ]$.
Performing a similarity transformation to trivialize the measure, we arrive at a manifestly hermitian Hamiltonian\cite{collft,Das:1990kaa}
\bea
H=\Pi \Omega \Pi+{1\over 4}\left(\omega+{\partial\Omega\over\partial\phi}\right)\Omega^{-1}
\left(\omega+{\partial\Omega\over\partial\phi}\right)+\phi_1-{1\over 2}{\partial\omega\over\partial\phi}
-{1\over 2}{\partial^2\Omega\over\partial\phi\partial\phi}
\eea
where we have used an obvious matrix notation and have introduced the momentum $\Pi(n)$ conjugate to $\phi_n$
\bea
\Pi_n=-i{\partial\over\partial\phi_n}
\eea
As we have commented above, the variables $\phi_n$ that we have employed in the description so far are a
natural generalization of the variables (\ref{traces}) used in the matrix model.
The variables (\ref{traces}) are essentially the eigenvalues of the matrix model, so that it is natural to
interpret the gauge invariant variables constructed out of $T^i{}_j $ as providing an eigenvalue like description
of the tensor model.
Just as in the matrix case, the quantum mechanical system develops an extra dimension.
To see how this happens, we can explore the range of the eigenvalues of $T$.
In the case of a single matrix, the change to eigenvalues induces a Van der Monde determinant which produces a
repulsion between the eigenvalues ensuring they spread out to produce a macroscopic emergent geometry at large $N$.
To get some insight into what is happening in the case of the tensor model, we compute the one point functions
\bea
\langle {\rm Tr}(T^k)\rangle=\langle\bar\Phi \cdot (k)\circ{\bf 1}\circ{\bf 1}\cdot\Phi\rangle
=\sum_{\sigma\in S_k}N_1^{C((k)\sigma)}N_2^{C(\sigma)}N_3^{C(\sigma)}
\eea
In the above $(k)$ is a $k$-cycle which, for concreteness, we take to be $(123\cdots k)$.
Lets study the limit that $N_i\to\infty$ holding ${N_2\over N_3}$ fixed and taking $\alpha={N_2 N_3\over N_1}$ fixed.
In this large $N_i$ limit the above sum is then dominated by a nontrivial class of diagrams.
For example
\bea
\langle {\rm Tr}(T)\rangle &=& N_1N_2N_3=\alpha N_1^2\cr
\langle {\rm Tr}(T^2)\rangle &=& N_1^2N_2N_3+N_1 N_2^2 N_3^2=(\alpha +\alpha^2) N_1^3\cr
\langle {\rm Tr}(T^3)\rangle &=& N_1^3N_2N_3+3N_1^2 N_2^2 N_3^2+N_1 N_2^3 N_3^3+N_1N_2N_3\cr
&=&(\alpha + 3\alpha^2 +\alpha^3+{\alpha\over N_1^2}) N_1^4\cr
\langle {\rm Tr}(T^4)\rangle &=& N_1^4 N_2 N_3+6N_1^3 N_2^2 N_3^2+6N_1^2 N_2^3 N_3^3+N_1 N_2^4 N_3^4
+5 N_1^2 N_2 N_3 + 5 N_1 N_2^2 N_3^2   \cr
&=&N_1^5\left( \alpha+6\alpha^2+6\alpha^3+\alpha^4
+{5\alpha\over N^2} + {5 \alpha^2\over N^2}\right)\cr
\langle {\rm Tr}(T^5)\rangle &=&
N_1^5 N_2 N_3+10 N_1^4 N_2^2 N_3^2+20 N_1^3 N_2^3 N_3^3+10 N_1^2 N_2^4 N_3^4+
N_1 N_2^5 N_3^5\cr
&&+15 N_1^3 N_2 N_3+40 N_1^2 N_2^2 N_3^2+15 N_1 N_2^3 N_3^3+8 N_1 N_2 N_3\cr
&=&
N_1^6\Big( \alpha+10 \alpha^2+20 \alpha^3+10 \alpha^4+\alpha^5
+{15 \alpha+40 \alpha^2+15\alpha^3\over N_1^2}+8 {\alpha\over N_1^4}\Big)\cr
\langle {\rm Tr}(T^k)\rangle &\sim& N_1^{k+1}\label{expdata}
\eea
The growth with $k$ as $N_1^{k+1}$ is a clear indication that the eigenvalues of $T$ are spreading out and are 
potentially able to generate a new dimension.
To construct the field theory in this extra dimension, it is useful to introduce the field
\bea
\phi (x)=\int {dk\over 2\pi}\, e^{-ikx}\, \phi_k\qquad
\phi_k={\rm Tr}(e^{ikT})
\eea
Notice that $\phi (x)$ is nothing but the density of eigenvalues of the $T$ matrix and consequently
\bea
{\rm Tr}(T^n)=\int dx\phi (x) x^n
\eea
The momentum dual to $\phi (x)$ is $\pi(x)={1\over i}{\delta\over\delta\phi (x)}$ and similarly
$\pi_k={1\over i}{\delta\over\delta\phi_k}$.
To perform the change of variables, note that the kinetic terms in the tensor model Hamiltonian can be written as
\bea
   -{\partial\over\partial\bar\phi^{ijk}}{\partial\over\partial\phi_{ijk}}&=&
-T^i{}_l {\partial\over\partial T^j{}_l} {\partial\over\partial T^i{}_j}
- N_2 N_3 {\partial\over\partial T^i{}_i}\cr
&=&-\int dk\int dk'\,\, \Omega_{k,k'}\,\pi_k\,\pi_{k'}
+\int dk\,\, \omega_k\,\pi_k
\eea
where
\bea
\Omega_{k,k'}&=&T^i{}_l {\partial\phi_{k}\over\partial T^j{}_l} {\partial\phi_{k'}\over\partial T^i{}_j}
=-kk' {\rm Tr}(T\, e^{i(k+k')T})\cr
&=&ikk'{\partial\over\partial k}\phi_{k+k'}
\eea
and
\bea
\omega_k&=&-T^i{}_l {\partial\over\partial T^j{}_l}\left( {\partial\phi_k\over\partial T^i{}_j}\right)
- N_2 N_3 {\partial\phi_k\over\partial T^i{}_i}\cr
&=&k\int_0^1 d\tau\,\, \phi_{\tau k}i{\partial\over\partial \tau}\phi_{(1-\tau )k}
-ikN_2 N_3\phi_{k}
\eea
To obtain these results, we have used the formula
\bea
{\partial\over\partial M^i{}_j}(e^{-ikM})^k_l=(-ik)\int_0^1 d\tau
(e^{-i\tau k M})^k{}_i (e^{-i(1-\tau)kM})^j{}_l
\eea
In position space we obtain
\bea
   \Omega (x,x')={\partial\over\partial x}{\partial\over\partial x'}\left( x\phi (x)\delta (x-x')\right)
\eea
and
\bea
\omega (x) =2{\partial\over\partial x}\dashint dy\,\, \phi(x)\phi(y){x\over x-y}
+(N_2 N_3-N_1){\partial\phi(x)\over\partial x}
\eea
It is interesting to note that the formula for $\Omega (x,x')$ is identical to the formula obtained from the
radial sector of multi matrix models, and that the formula for $\omega (x)$ is very similar 
- see  \cite{Masuku:2009qf,Masuku:2011pm,Masuku:2014wxa,Masuku:2015vta}.
This easily leads to the following Hamiltonian (we have dropped constant terms)
\bea
H&=&\int dx\left[ {\partial\pi\over\partial x}\, x\phi (x)\, {\partial\pi\over\partial x}
+{\phi (x)\over 4x}
\left(\dashint dy\,\, {2x\phi (y)\over x-y}\right)^2+{(N_2 N_3-N_1)^2\over 4x}\phi (x)+{x\over 4}\phi (x)\right]\cr
&&\qquad\qquad\qquad -\mu\int dx \, \phi (x)
\eea
where the last term above enforces the constraint $\int dx \phi (x)=N_1$.
To get this result, we used
\bea
\int dx\dashint dy\, \phi (x)\phi (y) {x+y\over x-y}=0
\eea
As we explain in Appendix \ref{collFTidentities}, the Hamiltonian can be written as
\bea
H=\int dx {\partial\pi\over\partial x}x\phi (x){\partial\pi\over\partial x} + V_{\rm eff}
\eea
where the effective potential is
\bea
V_{\rm eff}=\int dx\left[ {\pi^2 x\over 3}\phi^3
+{(\alpha -1)^2 N_1^2\over 4x}\phi (x)+{x\over 4}\phi (x)-\mu\phi(x)\right]
\eea
The classical field should minimize the effective potential, which leads to the following
classical collective equation of motion
\bea
0={\delta V_{\rm eff}\over\delta\phi (x)}=\pi^2 x\phi^2+{(1-\alpha)^2N_1^2\over 4x}
+{x\over 4}-\mu\cr\cr
\Rightarrow \phi(x)={1\over\pi}\sqrt{{\mu\over x}-{1\over 4}-{(1-\alpha)^2 N_1^2\over 4x^2}}
\eea
The chemical potential $\mu$ should be fixed by requiring that
\bea
\int_{x_-}^{x_+} dx\phi (x)= N_1
\eea
where the limits of integration are
\bea
x_{\pm}=2 \mu \pm \sqrt{4 \mu ^2-N_1^2(1-\alpha)^2}
\eea
As a test of this classical solution, we would like to show that it reproduces the correct large $N_1$ correlators.
To simplify the analysis that follows, we will set $\alpha=1$.
In this case, after solving for $\mu$ we have the density
\bea
\phi(x)={1\over\pi}\sqrt{{N_1\over x}-{1\over 4}}\qquad x_{+}=4N_1\quad x_-=0
\eea
A simple computation now gives
\bea
\int_0^{N_1\over 4} x \phi (x) dx = N_1^2&&\quad 
\int_0^{N_1\over 4} x^2\phi (x) dx =2 N_1^3\cr
\int_0^{N_1\over 4} x^3\phi (x) dx =5 N_1^4\quad &&
\int_0^{N_1\over 4} x^4\phi (x) dx =14 N_1^5\quad \int_0^{N_1\over 4} x^5\phi (x) dx =42 N_1^6
\eea
in complete agreement with (\ref{expdata}).
This provides a nice test of classical collective solution.

\section{Conclusions}

Motivated by the close connection of tensor models to the SYK model, we have considered the problem of counting
and then constructing the gauge invariant operators of tensor models.
Bosonic tensor models have already been considered in the literature, and the results we have obtained are 
consistent with what is already known.
Our results for fermionic vector models are novel.
Using the operators that have been constructed, we have exhibited an interesting algebra underlying the
gauge invariant operators of the tensor model: the gauge invariant operators define a ring.
We have written closed formulas for the structure constants of this ring.
As we have explained, this algebraic structure allows us to express arbitrary correlation functions as one
point functions, which we have computed explicitly.
Consequently, once the structure constants of the algebra are known, the free theory has been solved
exactly.
We have expressed these structure constants as overlaps of branching coefficients so that their computation
is now a well defined problem in the representation theory of the symmetric group.

To study the large $N$ dynamics of tensor model quantum mechanics we have identified a smaller set
of gauge invariant operators that has lead to an eigenvalue like description.
The system admits a collective field theory description which is similar but not identical to the
collective field theory of a singe hermitian matrix.
Our collective description shares all the good features of previous collective descriptions.
Two such features are
\begin{itemize}

\item[1.]
The collective description manifests the fact that the tensor model quantum mechanics has emergent dimensions.
Further, it is very attractive and highly non-trivial that the collective dynamics in this emergent dimension is local.

\item[2.]
The loop expansion parameter of the collective field theory is not $\hbar$ of the quantum mechanics, but rather
it is ${1\over N_1}$ with $N_1$ set by the tensor model gauge group.
Consequently the classical equations of motion of the collective field theory yield the answer obtained by summing the
complete set of Feynman diagrams that contribute at large $N_1$.
For our tensor model example we have explicitly demonstrated this. 

\end{itemize}
The above two features are highly suggestive of holography, which claims that a local (at large $N_1$)
higher dimensional classical system is dual to the large $N$ limit of the gauge theory.

There are a number of future directions that should be pursued.
The fermionic tensor model rather than the bosonic tensor model appears to be more relevant to the problem 
of understanding holography.
It would be interesting to develop the collective field theory of the fermionic model.
Specifically, it would be fascinating if such a description could be developed for the 
Witten-Gurau model, which is of most relevance for SYK.
Perhaps the most interesting question to ask is if we can enlarge the space of gauge invariants to get a genuinely 
larger space than the loop space of a single matrix model, such that the enlarged space is still manageable?
It seems that tensor models maybe good toy models with which to explore holography.

{\vskip 0.5cm}

\noindent
{\it Acknowledgements:}
We would like to thank Jo\~ao  Rodrigues for helpful discussions.
This work is based upon research supported by the South African Research Chairs
Initiative of the Department of Science and Technology and National Research Foundation.
Any opinion, findings and conclusions or recommendations expressed in this material
are those of the authors and therefore the NRF and DST do not accept any liability
with regard thereto.

\begin{appendix}

\section{Check of counting formulas}\label{CheckCount}

In this section we will explore the counting formulas obtained in Section \ref{CountConstruct}.
First consider the counting at infinite $N$.
It is rather easy to use characters of the symmetric group to compute the Kronecker coefficients
and then sum the squares of the coefficients, to compute the number of bosonic gauge invariant
operators ($N_b$) and the number of fermionic gauge invariants operators ($N_f$).
The results are shown in Table 1.

{\vskip 0.2cm}

\begin{table}[h]
\begin{center}
\begin{tabular}{|c|c|c|c|c|c|c|}
\hline
n & 1&2 &3 &4 &5&6\\
\hline
\hline
$N_b$ &1& 4 & 11 &43&161&901\\
\hline
$N_f$ &1& 4 & 11 &43&161&901\\
\hline
\end{tabular} \label{tab:BgN}
\end{center}
\caption{The number of bosonic $N_b$ or fermionic $N_f$ gauge invariant tensors constructed using $n$ fields.
This counting is for gauge group ranks $N_1=N_2=N_3=\infty$.}
\end{table}

{\vskip 0.2cm}

Note that the number of fermionic gauge invariant operators is equal to the number of bosonic gauge invariant operators. 
This fact is easily explained: every time we have a non-zero bosonic Kronecker coefficient, there is a corresponding
non-zero fermionic Kronecker coefficient.
This is easily proved using the well known property of characters of the symmetric group
\bea
\chi_{R^T}(\sigma)={\rm sgn}(\sigma)\chi_{R}(\sigma)
\eea
where $R^T$ is the transposed Young diagram, i.e. the Young diagram obtained from $R$ by swapping rows 
and columns.
For example
\bea
   R=\yng(3,2)\qquad\Rightarrow\qquad R^T=\yng(2,2,1)
\eea
Recall that $1^n$ represents the Young diagram with a single column of $n$ boxes. 
Use $n$ to denote the Young diagram that has a single row of $n$ boxes.
The proof is as follows 
\bea
g_{r_1,r_2,r_3,n}&=&{1\over n!}\sum_{\sigma\in S_n}
\chi_{r_1}(\sigma)\chi_{r_2}(\sigma)\chi_{r_3}(\sigma)\chi_{n}(\sigma)\cr
&=&{1\over n!}\sum_{\sigma\in S_n}
\chi_{r_1}(\sigma)\chi_{r_2}(\sigma)\chi_{r_3}(\sigma)\cr
&=&{1\over n!}\sum_{\sigma\in S_n}
\chi_{r_1^T}(\sigma)\chi_{r_2^T}(\sigma)\chi_{r_3^T}(\sigma) ({\rm sgn}(\sigma))^3\cr
&=&{1\over n!}\sum_{\sigma\in S_n}
\chi_{r_1^T}(\sigma)\chi_{r_2^T}(\sigma)\chi_{r_3^T}(\sigma) {\rm sgn}(\sigma)\cr
&=&{1\over n!}\sum_{\sigma\in S_n}
\chi_{r_1^T}(\sigma)\chi_{r_2^T}(\sigma)\chi_{r_3^T}(\sigma)\chi_{1^n}(\sigma)\cr
&=&g_{r_1^T,r_2^T,r_3^T,1^n}
\eea
Since the set of non-zero Kronecker coefficients are the same for the bosonic and the fermionic tensor models, and the
number of gauge invariant operators is equal to the sum of the squares of these coefficients, this proves that the
number of gauge invariant operators one can construct in bosonic tensor models equals the number of gauge invariant
operators one can construct in fermionic tensor models.

The argument above has been for rank $d=3$ tensors.
It is clear that the above proof goes through for rank $d$ tensors with $d$ odd, since in this case
$({\rm sgn}(\sigma))^d={\rm sgn}(\sigma)$.
For even $d$ however, the above proof does not go through: in this case
$({\rm sgn}(\sigma))^d=1$.
However, a simple variant of the proof does work: for rank four for example, it is simple to prove that
\bea
g_{r_1,r_2,r_3,r_4,n}=g_{r_1^T,r_2^T,r_3^T,r_4,1^n}
\eea
We have verified this equality explicitly for $n\le 6$ and ranks $d\le 8$, which is a nice check of the above arguments.  

At finite $N$ the number of fermionic and bosonic gauge invariant operators no longer matches.
Recall that for a general rank $d$ tensor model the gauge group is $U(N_1)\times U(N_2)\times \cdots\times U(N_d)$.
As soon as $n$ exceeds any of the $N_i$, it is possible to have Young diagrams $r\vdash n$ whose number of rows
is greater than at least one of the $N_i$.
In this case, the corresponding operator vanishes and it is for this reason that we must put a cut off on the number of
rows.
For example, for the bosons we have
\bea
N_b=\sum_{r_i\vdash n\,\, l(r_i)\le N_i}g_{r_1\, r_2\, r_3\, 1}^2
\eea
The proof breaks because we can have, for example, $l(r_1)< N_1$ and $l(r_1^T)> N_1$.
In Table 2 we have given the finite $N$ counting for rank $3$ tensors with $N_1=N_2=N_3=5$.

{\vskip 0.2cm}

\begin{table}[h]
\begin{center}
\begin{tabular}{|c|c|c|c|c|c|c|}
\hline
n & 1&2 &3 &4 &5&6\\
\hline
\hline
$N_b$ &1& 4 & 11 &43&92&70\\
\hline
$N_f$ &1& 4 & 11 &43&87&20\\
\hline
\end{tabular} \label{Nis5}
\end{center}
\caption{The number of bosonic $N_b$ or fermionic $N_f$ gauge invariant tensors constructed using $n$ fields.
Here $N_1=N_2=N_3=5$.}
\end{table}

{\vskip 0.2cm}

Notice that there are more bosonic gauge invariant operators than there are fermionic gauge invariant operators.
This is in fact rather general: the Kronecker coefficients relevant for the bosonic gauge invariants are mostly
short and wide Young diagrams.
On the other hand, the Kronecker coefficients relevant for the fermionic gauge invariants are mostly
tall and thin Young diagrams.
In fact, for the fermionic tensor model, there is some value of $n$ beyond which there are no new gauge invariants.
In Table 3 we have shown the finite $N$ counting for rank $3$ tensors with $N_1=N_2=N_3=3$.
For $n\ge 6$ there are no gauge invariant operators.

{\vskip 0.2cm}

\begin{table}[h]
\begin{center}
\begin{tabular}{|c|c|c|c|c|c|c|}
\hline
n & 1&2 &3 &4 &5&6\\
\hline
\hline
$N_b$ &1& 4 & 11 &12&151&18\\
$N_f$ &1& 4 & 11 &8   &  41&0\\
\hline
\end{tabular}
\end{center} 
\caption{The number of bosonic $N_b$ or fermionic $N_f$ gauge invariant tensors constructed using $n$ fields.
Here $N_1=N_2=N_3=3$.}
\end{table}

{\vskip 0.2cm}

For the first few values of $n$, it is possible to explicitly construct the gauge invariant operators. 
For $n=1$ there is a single bosonic and a single fermionic gauge invariant operator
\bea
\bar\phi^{ijk}\phi_{ijk}\qquad\qquad \bar\psi^{ijk}\psi_{ijk}
\eea
For $n=2$ we have the following bosonic operators
\bea
\bar\phi^{i_1 j_1 k_1}\bar\phi^{i_2j_2k_2}\phi_{i_1 j_1 k_1}\phi_{i_2j_2k_2}\qquad\qquad 
\bar\phi^{i_2 j_1 k_1}\bar\phi^{i_1j_2k_2}\phi_{i_1 j_1 k_1}\phi_{i_2j_2k_2}\cr
\bar\phi^{i_1 j_2 k_1}\bar\phi^{i_2j_1k_2}\phi_{i_1 j_1 k_1}\phi_{i_2j_2k_2}\qquad\qquad 
\bar\phi^{i_1 j_1 k_2}\bar\phi^{i_2j_2k_1}\phi_{i_1 j_1 k_1}\phi_{i_2j_2k_2}
\eea
which nicely matches the counting given above.
There is an identical set of operators for the fermions.
For $n=3$ we have the following bosonic operators
\be 
\begin{array}{lll}
\bar\phi^{i_1j_1k_1}\bar\phi^{i_2j_2k_2}\bar\phi^{i_3j_3k_3}\phi_{i_1j_1k_1}\phi_{i_2j_2k_2}\phi_{i_3j_3k_3} &\;& \bar\phi^{i_1j_1k_1}\bar\phi^{i_2j_2k_3}\bar\phi^{i_3j_3k_2}\phi_{i_1j_1k_1}\phi_{i_2j_2k_2}\phi_{i_3j_3k_3}\\
\bar\phi^{i_1j_1k_1}\bar\phi^{i_2j_3k_2}\bar\phi^{i_3j_3k_2}\phi_{i_1j_1k_1}\phi_{i_2j_2k_2}\phi_{i_3j_3k_3} &\;& \bar\phi^{i_1j_1k_1}\bar\phi^{i_2j_3k_3}\bar\phi^{i_3j_2k_2}\phi_{i_1j_1k_1}\phi_{i_2j_2k_2}\phi_{i_3j_3k_3} \\
\bar\phi^{i_1j_1k_2}\bar\phi^{i_2j_2k_3}\bar\phi^{i_3j_3k_1}\phi_{i_1j_1k_1}\phi_{i_2j_2k_2}\phi_{i_3j_3k_3} &\; &\bar\phi^{i_1j_1k_2}\bar\phi^{i_2j_3k_1}\bar\phi^{i_3j_2k_3}\phi_{i_1j_1k_1}\phi_{i_2j_2k_2}\phi_{i_3j_3k_3}\\
\bar\phi^{i_1j_1k_2}\bar\phi^{i_2j_3k_3}\bar\phi^{i_3j_2k_1}\phi_{i_1j_1k_1}\phi_{i_2j_2k_2}\phi_{i_3j_3k_3} &\; & \bar\phi^{i_1j_2k_1}\bar\phi^{i_2j_3k_2}\bar\phi^{i_3j_1k_3}\phi_{i_1j_1k_1}\phi_{i_2j_2k_2}\phi_{i_3j_3k_3}\\
\bar\phi^{i_1j_2k_1}\bar\phi^{i_2j_3k_3}\bar\phi^{i_3j_1k_2}\phi_{i_1j_1k_1}\phi_{i_2j_2k_2}\phi_{i_3j_3k_3} &\;& \bar\phi^{i_1j_2k_2}\bar\phi^{i_2j_3k_3}\bar\phi^{i_3j_1k_1}\phi_{i_1j_1k_1}\phi_{i_2j_2k_2}\phi_{i_3j_3k_3}\\
\bar\phi^{i_1j_2k_3}\bar\phi^{i_2j_3k_1}\bar\phi^{i_3j_1k_2}\phi_{i_1j_1k_1}\phi_{i_2j_2k_2}\phi_{i_3j_3k_3}
\end{array}
\ee
which matches the counting given above. The set of fermionic operators is again the same.
\section{Examples of Operators and Correlators}\label{Ops}

In the previous Appendix we have written down some of the gauge invariant operators in the permutation basis.
In this Appendix we will write down some operators in the representation theory basis. 
We will then explore correlators of gauge invariant operators, in both bases.

For $n=2$ fields, there are no multiplicities, so these labels are dropped.
There is a total of four gauge invariant operators that can be defined.
We will give the complete set of gauge invariant operators, since this will allow us to test that they are
indeed orthogonal and have the correct two point function.
The operators are given by
\bea
{\cal O}_{\tiny\yng(2),\yng(2),\yng(2)}&=&
2\bar\phi^{i_1 j_1 k_1}\bar\phi^{i_2j_2k_2}\phi_{i_1 j_1 k_1}\phi_{i_2j_2k_2}+ 
2\bar\phi^{i_2 j_1 k_1}\bar\phi^{i_1j_2k_2}\phi_{i_1 j_1 k_1}\phi_{i_2j_2k_2}\cr
&&+2\bar\phi^{i_1 j_2 k_1}\bar\phi^{i_2j_1k_2}\phi_{i_1 j_1 k_1}\phi_{i_2j_2k_2}
+2\bar\phi^{i_1 j_1 k_2}\bar\phi^{i_2j_2k_1}\phi_{i_1 j_1 k_1}\phi_{i_2j_2k_2}
\eea
\bea
{\cal O}_{\tiny\yng(1,1),\yng(1,1),\yng(2)}&=&
2\bar\phi^{i_1 j_1 k_1}\bar\phi^{i_2j_2k_2}\phi_{i_1 j_1 k_1}\phi_{i_2j_2k_2}-
2\bar\phi^{i_2 j_1 k_1}\bar\phi^{i_1j_2k_2}\phi_{i_1 j_1 k_1}\phi_{i_2j_2k_2}\cr
&&-2\bar\phi^{i_1 j_2 k_1}\bar\phi^{i_2j_1k_2}\phi_{i_1 j_1 k_1}\phi_{i_2j_2k_2}
+2\bar\phi^{i_1 j_1 k_2}\bar\phi^{i_2j_2k_1}\phi_{i_1 j_1 k_1}\phi_{i_2j_2k_2}
\eea
\bea
{\cal O}_{\tiny\yng(1,1),\yng(2),\yng(1,1)}&=&
2\bar\phi^{i_1 j_1 k_1}\bar\phi^{i_2j_2k_2}\phi_{i_1 j_1 k_1}\phi_{i_2j_2k_2}-
2\bar\phi^{i_2 j_1 k_1}\bar\phi^{i_1j_2k_2}\phi_{i_1 j_1 k_1}\phi_{i_2j_2k_2}\cr
&&+2\bar\phi^{i_1 j_2 k_1}\bar\phi^{i_2j_1k_2}\phi_{i_1 j_1 k_1}\phi_{i_2j_2k_2}
-2\bar\phi^{i_1 j_1 k_2}\bar\phi^{i_2j_2k_1}\phi_{i_1 j_1 k_1}\phi_{i_2j_2k_2}
\eea
\bea
{\cal O}_{\tiny\yng(2),\yng(1,1),\yng(1,1)}&=&
2\bar\phi^{i_1 j_1 k_1}\bar\phi^{i_2j_2k_2}\phi_{i_1 j_1 k_1}\phi_{i_2j_2k_2}+ 
2\bar\phi^{i_2 j_1 k_1}\bar\phi^{i_1j_2k_2}\phi_{i_1 j_1 k_1}\phi_{i_2j_2k_2}\cr
&&-2\bar\phi^{i_1 j_2 k_1}\bar\phi^{i_2j_1k_2}\phi_{i_1 j_1 k_1}\phi_{i_2j_2k_2}
-2\bar\phi^{i_1 j_1 k_2}\bar\phi^{i_2j_2k_1}\phi_{i_1 j_1 k_1}\phi_{i_2j_2k_2}
\eea
A simple but tedious computation confirms (\ref{bos1pnt}) and (\ref{bos2pnt}).
Some sample computations are
\bea
\langle {\cal O}_{\tiny\yng(2),\yng(2),\yng(2)} \rangle=2N_1(N_1+1)N_2(N_2+1)N_3(N_3+1)
\eea
\bea
\langle {\cal O}_{\tiny\yng(2),\yng(2),\yng(2)} {\cal O}_{\tiny\yng(2),\yng(2),\yng(2)} \rangle =
32N_1(N_1+1)N_2(N_2+1)N_3(N_3+1)
\eea
\bea
\langle {\cal O}_{\tiny\yng(2),\yng(2),\yng(2)} {\cal O}_{\tiny\yng(2),\yng(2),\yng(1,1)}\rangle
=0
\eea
\\
For $n=2$, the complete set of fermionic operators in the representation basis is given by 
\bea 
\mathcal{P}_{\tiny\yng(1,1),\yng(1,1),\yng(1,1)}&=&2\bar\psi^{i_1j_1k_1}\bar\psi^{i_2j_2k_2}\psi_{i_2j_2k_2}\psi_{i_1j_1k_1}-2\bar\psi^{i_1j_2k_1}\bar\psi^{i_2j_1k_2}\psi_{i_2j_2k_2}\psi_{i_1j_1k_1} \cr
&&-2\bar\psi^{i_1j_1k_2}\bar\psi^{i_2j_2k_1}\psi_{i_2j_2k_2}\psi_{i_1j_1k_1}-2\bar\psi^{i_2j_1k_1}\bar\psi^{i_1j_2k_2}\psi_{i_2j_2k_2}\psi_{i_1j_1k_1}
\eea
\bea 
\mathcal{P}_{\tiny\yng(1,1),\yng(2),\yng(2)}&=&2\bar\psi^{i_1j_1k_1}\bar\psi^{i_2j_2k_2}\psi_{i_2j_2k_2}\psi_{i_1j_1k_1}-2\bar\psi^{i_2j_1k_1}\bar\psi^{i_1j_2k_2}\psi_{i_2j_2k_2}\psi_{i_1j_1k_1} \cr
&&+2\bar\psi^{i_1j_2k_1}\bar\psi^{i_2j_1k_2}\psi_{i_2j_2k_2}\psi_{i_1j_1k_1}+2\bar\psi^{i_1j_1k_2}\bar\psi^{i_2j_2k_1}\psi_{i_2j_2k_2}\psi_{i_1j_1k_1}
\eea
\bea 
\mathcal{P}_{\tiny\yng(2),\yng(1,1),\yng(2)}&=&2\bar\psi^{i_1j_1k_1}\bar\psi^{i_2j_2k_2}\psi_{i_2j_2k_2}\psi_{i_1j_1k_1}+2\bar\psi^{i_2j_1k_1}\bar\psi^{i_1j_2k_2}\psi_{i_2j_2k_2}\psi_{i_1j_1k_1} \cr
&&-2\bar\psi^{i_1j_2k_1}\bar\psi^{i_2j_1k_2}\psi_{i_2j_2k_2}\psi_{i_1j_1k_1}+2\bar\psi^{i_1j_1k_2}\bar\psi^{i_2j_2k_1}\psi_{i_2j_2k_2}\psi_{i_1j_1k_1}
\eea
\bea 
\mathcal{P}_{\tiny\yng(2),\yng(2),\yng(1,1)}&=&2\bar\psi^{i_1j_1k_1}\bar\psi^{i_2j_2k_2}\psi_{i_2j_2k_2}\psi_{i_1j_1k_1}+2\bar\psi^{i_2j_1k_1}\bar\psi^{i_1j_2k_2}\psi_{i_2j_2k_2}\psi_{i_1j_1k_1} \cr
&&+2\bar\psi^{i_1j_2k_1}\bar\psi^{i_2j_1k_2}\psi_{i_2j_2k_2}\psi_{i_1j_1k_1}-2\bar\psi^{i_1j_1k_2}\bar\psi^{i_2j_2k_1}\psi_{i_2j_2k_2}\psi_{i_1j_1k_1}
\eea
Some sample computations confirming (\ref{fermion1pt}) and (\ref{fermion2pt}) are
\bea
\langle \mathcal{P}_{\tiny\yng(1,1),\yng(1,1),\yng(1,1)}\rangle=2N_1(N_1-1)N_2(N_2-1)N_3(N_3-1)
\eea
\bea
\langle \mathcal{P}_{\tiny\yng(1,1),\yng(1,1),\yng(1,1)}\mathcal{P}_{\tiny\yng(1,1),\yng(1,1),\yng(1,1)}\rangle=32N_1(N_1-1)N_2(N_2-1)N_3(N_3-1)
\eea
\bea
\langle \mathcal{P}_{\tiny\yng(1,1),\yng(1,1),\yng(1,1)}\mathcal{P}_{\tiny\yng(1,1),\yng(1,1),\yng(2)}\rangle=0
\eea

\section{Identities needed to derive the collective field theory Hamiltonian}\label{collFTidentities}

Using the identity
\bea
   \dashint dy {e^{-iky}\over x-y}=\epsilon (k)\pi i e^{-ikx}
\eea
we find
\bea
   \dashint dy\,\, {2x\,e^{-iky}\over y-x}=2x\epsilon (k)\pi i e^{-ikx}
\eea
Our main goal in this Appendix is to explain how to rewrite the term
\bea
T_1=\int dx {\phi (x)\over x}
\dashint dy_1 {2x\phi (y_1)\over y_1-x}\dashint dy_2 {2x \phi (y_2)\over y_2-x}
\eea
in a manifestly local form.
This is the only term in the Hamiltonian that is not manifestly local.
Use the Fourier transform
\bea
\phi (x)=\int{dk\over 2\pi} e^{-ikx}\phi_k
\eea 
to write (this is the only non-local term in the Hamiltonian)
\bea
T_1&=&\int {dk_1\over 2\pi}\int {dk_2\over 2\pi}\int {dk_3\over 2\pi}\int dx
\phi_{k_1}\phi_{k_2}\phi_{k_3}4x e^{-i k_1 x}
\dashint dy_1 {e^{-ik_2y_1}\over y_1-x}
\dashint dy_2 {e^{-ik_3y_2}\over y_2-x}\cr
&&=\int {dk_1\over 2\pi}\int {dk_2\over 2\pi}\int {dk_3\over 2\pi}\int dx
\phi_{k_1}\phi_{k_2}\phi_{k_3}4x e^{-i k_1 x}\left[\pi i\epsilon (k_2)e^{-ik_2 x}\right]
\left[\pi i\epsilon (k_3)e^{-ik_3 x}\right]\cr
&=&
-4\pi^2\int {dk_1\over 2\pi}\int {dk_2\over 2\pi}\int {dk_3\over 2\pi}\int dx
\phi_{k_1}\phi_{k_2}\phi_{k_3}x e^{-i (k_1+k_2+k_3) x}\epsilon (k_2)\epsilon (k_3)
\eea
The expression on the last line can be manipulated, by renaming variables into
\bea
&&-{4\pi^2\over 3}\int {dk_1\over 2\pi}\int {dk_2\over 2\pi}\int {dk_3\over 2\pi}\int dx
\phi_{k_1}\phi_{k_2}\phi_{k_3}x e^{-i (k_1+k_2+k_3) x}
(\epsilon (k_1)\epsilon (k_2)+\epsilon (k_1)\epsilon (k_3)+\epsilon (k_2)\epsilon (k_3))\cr
&&={4\pi^2\over 3}\int {dk_1\over 2\pi}\int {dk_2\over 2\pi}\int {dk_3\over 2\pi}
\phi_{k_1}\phi_{k_2}\phi_{k_3} (i\partial_{k_1}\delta (k_1+k_2+k_3))
(\epsilon (k_1)\epsilon (k_2)+\epsilon (k_1)\epsilon (k_3)+\epsilon (k_2)\epsilon (k_3))\cr
&&
\eea
Because of the delta function, one or two of the $k_i$'s must be positive so that 
\bea
\epsilon (k_1)\epsilon (k_2)+\epsilon (k_1)\epsilon (k_3)+\epsilon (k_2)\epsilon (k_3)=-1
\eea
and we now find
\bea
T_1={4\pi^2\over 3}\int {dk_1\over 2\pi}\int {dk_2\over 2\pi}\int {dk_3\over 2\pi}
\phi_{k_1}\phi_{k_2}\phi_{k_3} (i\partial_{k_1}\delta (k_1+k_2+k_3))
={4\pi^2\over 3}\int dx x\phi^3 (x)
\eea
so that, remarkably, this term is local and it gives rise to a cubic interaction!

\end{appendix}

\end{document}